\documentclass{aa} 
\usepackage{graphicx} 
\usepackage[varg]{txfonts}
\usepackage{booktabs}
\usepackage{longtable}
\usepackage{natbib}
\bibpunct{(}{)}{;}{a}{}{,}
\usepackage[dvipsnames]{xcolor}
\usepackage{placeins}

\makeatletter
\renewcommand*\aa@pageof{, page \thepage{} of \pageref*{LastPage}}
\makeatother

\begin{document}

\title{High-cadence stellar variability studies of 
RR Lyrae stars\\ with DECam: New multi-band templates}
\titlerunning{$griz$ RR Lyrae templates} \authorrunning{Baeza-Villagra et al.} 

\author{K. Baeza-Villagra\inst{\ref{inst1},\ref{inst2},\ref{inst3},\ref{inst4},\ref{inst5}}\orcid{0000-0003-2804-1261}
\and N. Rodr{\'i}guez-Segovia\inst{\ref{inst6}}\orcid{0000-0002-0125-1472}
\and M. Catelan\inst{\ref{inst1},\ref{inst2}}\orcid{0000-0001-6003-8877} 
\and A. Rest\inst{\ref{inst7},\ref{inst8}}\orcid{0000-0002-4410-5387}
\and A. Papageorgiou\inst{\ref{inst9}}\orcid{0000-0002-3039-9257}
\and C. E. Mart\'{i}nez-V\'{a}zquez\inst{\ref{inst10}}\orcid{0000-0002-9144-7726}
\and A. A. R. Valcarce\inst{\ref{inst11}}\orcid{0000-0003-4623-3961}
\and C. E. Ferreira Lopes\inst{\ref{inst12},\ref{inst2}}\orcid{0000-0002-8525-7977}
\and F. B. Bianco\inst{\ref{inst13},\ref{inst14},\ref{inst15},\ref{inst16}}\orcid{}
}

\institute{Instituto de Astrof\'isica, Pontificia Universidad Cat\'olica de Chile, Av. Vicu\~na Mackenna 4860, 7820436 Macul, Santiago, Chile 
\label{inst1}
\and Millennium Institute of Astrophysics, Nuncio Monse{\~n}or S{\'o}tero Sanz 100, Providencia, Santiago, Chile\label{inst2}
\and Dipartimento di Fisica, Università di Roma Tor Vergata, Via della Ricerca Scientifica, 1, Roma 00133, Italy\label{inst3} 
\and Dipartimento di Fisica, Sapienza Università di Roma, P.le A. Moro 5, Roma 00185, Italy\label{inst4} 
\and INAF—Osservatorio Astronomico di Roma, via Frascati 33, I-00078 Monte Porzio Catone, Italy \label{inst5}
\and School of Science, University of New South Wales, Australian Defence Force Academy, Canberra, ACT 2600, Australia\label{inst6}
\and Space Telescope Science Institute, 3700 San Martin Drive, Baltimore, MD 21218, USA\label{inst7}
\and 
Department of Physics and Astronomy, The Johns Hopkins University, 366 Bloomberg Center, 3400 N. Charles Street, Baltimore, MD 21218, USA\label{inst8}
\and Department of Physics, University of Patras, 26500, Patra, Greece\label{inst9}
\and International Gemini Observatory/NSF NOIRLab, 670 N. A'ohoku Place, Hilo, Hawai'i, 96720, USA\label{inst10}
\and Departamento de Física, FACI, Universidad de Tarapacá, Casilla 7D, Arica, Chile\label{inst11}
\and  Instituto de Astronom\'{i}a y Ciencias Planetarias, Universidad de Atacama, Copayapu 485, Copiap\'{o} Chile\label{inst12}
\and
University of Delaware, Department of Physics and Astronomy, 217 Sharp Lab, Newark, DE 19716, USA\label{inst13}
\and
University of Delaware, Joseph R. Biden, Jr. School of Public Policy and Administration, 184 Academy Street, Newark, DE 19716, USA\label{inst14}
\and
University of Delaware, Data Science Institute, Newark, DE 19716, USA\label{inst15}
\and
Vera C. Rubin Observatory, Tucson, AZ 85719, USA\label{inst16}
} 

\date{Received /
Accepted }
\abstract{We present the most extensive set to date of high-quality RR Lyrae light curve templates in the $griz$ bands, based on time-series observations of the 
Dark Energy Camera Plane Survey
(DECaPS) East field, located in the Galactic bulge at coordinates (RA, DEC)(J2000) = (18:03:34, -29:32:02),  
obtained with the Dark Energy Camera (DECam) on the 4-m Blanco telescope at the Cerro Tololo Inter-American Observatory (CTIO). 
Our templates, which cover both fundamental-mode (RRab) and first-overtone (RRc) pulsators, can be especially useful when there is insufficient data for accurately calculating the average magnitudes and colors, hence distances, as well as to inform multi-band light curve classifiers, as will be required in the case of the Vera C. Rubin Observatory’s Legacy Survey of Space and Time (LSST).
In this paper, we describe in detail the procedures that were adopted in producing these templates, including a novel approach to account for the presence of outliers in photometry. Our final sample comprises 136 RRab and 144 RRc templates, all of which are publicly available.  
Lastly, in this paper we study the inferred Fourier parameters and other light curve descriptors, including rise time, skewness, and kurtosis, as well as their correlations with the pulsation mode, period, and effective wavelength.}


\maketitle 

\section{Introduction}
Pulsating stars comprise a subset of variable stars whose intrinsic physical properties change more or less cyclically over time. They are characterized by observable brightness changes over timescales ranging from minutes to years. This makes them a particularly interesting group of stars, as the variations in their physical properties happen in human timescales \citep[for a comprehensive overview, see, e.g.,][]{percy2007,PS}, and can therefore be studied in depth by observational campaigns that can cover multiple pulsation cycles. Amongst these stars, one particular subset is the RR Lyrae class (RRL, hereafter), whose members are a natural product of the evolution of low-mass stars \citep[0.55 to 0.8 $M_\odot$; e.g.,][]{PS,Marconi2015} that eventually reach the horizontal branch (HB) after helium burning starts in their core. It is during this stage that they cross the instability strip (IS) region of the Hertzsprung-Russell (HR) diagram and begin to pulsate regularly, with periods ranging from $\sim$ 0.2 to 1.0 days \citep{PS}, temperatures restricted to the range 6000~-- 7250 K, and radii between 4 to 6 $R_\odot$ \citep{Marcio2004,Marconi2005}. An additional consequence of their evolutionary history as low-mass stars is that they are usually associated with stellar populations older than $\sim$10 Gyr \citep{Walker1989,1995CAS....27.....S,Catelan2009,Savino2020}. Depending on their pulsation mode, RRLs can be divided into four sub-types \citep{1995CAS....27.....S}: RRab (fundamental mode), RRc (first-overtone), RRd (double-mode), and RRe \citep[second-overtone, which however have also been interpreted as first-overtone pulsators with exceptionally short periods; see e.g.,][]{1998ASPC..135...52K,2001AJ....122.2587C,2004ASPC..310..113C}. 

Henrietta Leavitt's pioneering work in 1912 \citep{1912Leavitt} established the foundation for using pulsating stars as distance indicators by discovering the period-luminosity (PL) relationship followed by classical Cepheids. Historically, in addition to Cepheids, other classes of pulsating stars, including RRL, have been utilized for this purpose with their own period-luminosity-metallicity relationships \citep[see, e.g., the recent volume by][]{deGrijs2024}. 
Considering the tight correlation between their pulsation periods and intrinsic luminosities, especially in the near-infrared, RRLs stand out as crucial stellar candles for measuring distances to old stellar populations \citep{Marcio2004, Braga2015, Saha_2019}.

In light of the extensive datasets generated by large-scale time-domain surveys and facilities like  the Panoramic Survey Telescope and Rapid Response System 
\citep[Pan-STARRS;][]{2017AJ....153..204S}, the Optical Gravitational Lensing Experiment \citep[OGLE;][]{2019AcA....69..321S}, the All-Sky Automated Survey for SuperNovae \citep[ASAS-SN;][]{2019MNRAS.486.1907J}, the {\em Gaia} Mission \citep{2018A&A...618A..30H,2019A&A...622A..60C,2022yCat..36740018C}, and others \citep[for a
review, see, e.g.,][]{Marcio2023}, the field of  variable star research has experienced some important advances in recent years. The forthcoming Vera C. Rubin Observatory's Legacy Survey of Space and Time (LSST) \citep{2022arXiv220804499H,2023arXiv230617333U} is anticipated to augment these endeavors by contributing a wealth of observational data not only on variable stars, but also transient events, active galactic nuclei, and other sources. The sheer volume of data will require sophisticated automatic classification and analysis tools that do not rely on spectroscopy. In this context, the employment of  multi-band template light curves is crucial for the development of new photometric classifiers
and for the proper classification and characterization of these stars.

In recent decades, significant advancements have been made in constructing light-curve templates for RRLs. \citet{1998AJ....115..193L} pioneered the creation of 6 Johnson V-band RRab light curve templates. These templates were utilized to estimate these stars' average magnitude and luminosity amplitude simultaneously. Later, \citet{sesar} developed optical multi-band templates for RRLs using Sloan Digital Sky Survey (SDSS) photometry in $ugriz$ bands. This comprehensive effort resulted in 22 RRab templates and 2 RRc templates, facilitating RRL identification \citep{2013AJ....146...94B,2017ApJ...834..160N,SesarHernitschek2017,vivas2017,2023MNRAS.519.5689M}. More recently, \citet{2019A&A...625A...1B} provided near-infrared JHKs light-curve templates for RRL variables.

We aim to provide new $griz$ multi-band templates for RRLs in the Dark Energy Camera Plane Survey (DECaPS) East (DCP-E) field \citep{2023MNRAS.519.3881G}. This field covers a 2.2 square degree area as determined by the field of view of the Dark Energy Camera (DECam, instrument through which it is periodically observed), and is located in the Galactic bulge at central coordinates (RA, DEC) (J2000) = (18:03:34, -29:32:02). These new templates will prove invaluable for next-generation multi-band classifiers used in extensive photometric surveys and could potentially improve the classification accuracy of currently available systems such as the Automatic Learning for the Rapid Classification of Events \citep[ALeRCE;][]{ALERCE2021,ALERCE2023}, as well as the accurate computation of RR Lyrae mean magnitudes and colors.

The paper is organized as follows. Section~\ref{sec:2} describes the data and methods used. The $griz$ multi-band templates are presented in Section~\ref{sec:3}. Results and their discussion are given and discussed in Sections~\ref{sec:4} and ~\ref{sec:5}, respectively. Finally, we present our conclusions in Section~\ref{sec:6}.

\section{Data and methods}\label{sec:2}

\subsection{Field selection and exposure times}
Time-domain astronomy has been historically challenged by working with irregularly spaced data due to solar motion, weather patterns, and unpredictable telescope allocations, resulting in wide gaps in their observations \citep{2019arXiv190108009F}. However, the large amount of high-quality data and modern classification methods offer an unprecedented opportunity to better characterize, classify, and understand variable and transient phenomena.

The DECam Alliance for Transients (DECAT) is a consortium of DECam Principal Investigators (PIs) involved in time-domain research, each typically requiring only a few hours per night for their respective projects. They collaborate by requesting co-scheduled shared full or half nights and collectively develop observation plans that incorporate targets from all programs \citep{Graham2021}. In this framework, time-series surveys of select deep-drilling fields (DDFs) are being conducted using DECam on the Blanco 4-m telescope at Cerro Tololo Inter-American Observatory (CTIO). Additionally, our team carried out two nights of high-cadence (June 2 and 3, 2021), multi-band ($griz$) observations of the DCP-E field during semester 2021A, under National Optical-Infrared Astronomy Research Laboratory (NOIRLab) proposal number 2021A-0921 (principal investigator: M. Catelan). The selection of the DCP-E field was based on maximizing the potential number of variable stars, as it partially overlaps the RR Lyrae-rich field B1 from \citet{Saha_2019}. B1 is centered on Baade’s Window, a region of relatively low obscuration that is close to the direction of the Galactic center \citep{Baade1951, Arp1965}.

Table~\ref{tab: number} presents a list of the acquired images from the DCP-E field for the regular DDF program  \citep{2023MNRAS.519.3881G} and our two nights of observations, categorized by filters. This work utilizes only images obtained up to May 2022.

\begin{table}[ht]
\centering
\caption{Number of images by filter obtained in DCP-E field for our two nights of observations (2021A-0921) and the DDF regular program \citep{2023MNRAS.519.3881G}.}
\label{tab: number}
\begin{tabular}{lcccccc}
\hline\hline
Program & \multicolumn{5}{c}{--------- Number of Images ---------} \\ 
 & $g$ & $r$ & $i$ & $z$ & Total & \\ 
\hline
DDF regular program  & 239 & 249 & 257 & 271 & 1016  \\
2021A-0921 (June 2) &2 & 437 & 0 & 0 & 439 \\
2021A-0921 (June 3) &114&115&114& 114& 457 \\
\hline
Total &355&801&371&385&1912 \\
\hline
\end{tabular}
\end{table}
Our own observations were carried out as follows. 

On the first night, we utilized $gr$ sequences with exposure times of 96 seconds for $g$ and 30 seconds for $r$. On the second night, we employed $griz$ sequences with exposure times of 96 seconds for $g$ and 30 seconds for the remaining filters. We did not use the $u$ and $Y$ exposures since they would not reach the desired depth compared to the $griz$ exposures, while at the same time impairing the desired high cadence. 
To ensure high-quality photometry, the observations presented in this paper were all conducted at airmasses lower than 2.

\subsection{Data processing}
\subsection{Photometry}
We processed the \texttt{instcal} images acquired from the NOIRLab archive\footnote{\href{https://astroarchive.noirlab.edu}{https://astroarchive.noirlab.edu}} by using a modified implementation of the \texttt{photpipe} pipeline designed explicitly for DECam images. \texttt{Photpipe} is a widely-used pipeline in various time-domain surveys \citep[e.g.,][]{2005ApJ...634.1103R,2007ApJ...666..674M,2014ApJ...795...44R,2016ApJ...826L..29C}. It is designed to perform single-epoch image processing tasks such as image calibration, photometric and astrometric calibration, image deformation, and obtain point-spread function (PSF) photometry using \textsc{dophot} \citep{Schechter_1993, AG2012}. 

The final catalogs produced by \texttt{photpipe} contain, amongst other properties, the coordinates, PSF flux, zero points, and \textsc{dophot} image type (\texttt{DoType}) quality flag for each detected source within a given image. We used sources with \texttt{DoType} = 1, ensuring that they are stellar sources with successful PSF fits \citep[for further information, see][]{DoPHOT}. Besides this requirement, we also decided to incorporate data from the regular DDF program conducted at the DCP-E field \citep{2023MNRAS.519.3881G} to maximize the phase coverage of stars within our sample. Importantly, both sets of images underwent identical processing to ensure uniformity. 

\subsection{Building the source catalog}
The literature was thoroughly reviewed to identify confirmed and candidate variable stars in the DCP-E field. We chose to analyze the RRLs listed in the OGLE catalog due to the exceptional quality of its light curves \citep{2014AcA....64..177S, 2019AcA....69..321S}. The OGLE database offers comprehensive and well-characterized light curves for numerous RRLs, enabling accurate and detailed analysis of their pulsation behavior. Moreover, it contains the largest number of RRLs among the catalogs we reviewed, ensuring a robust and extensive dataset for our study. We utilize the periods provided by the OGLE Collection of Variable Stars\footnote{\href{https://ogledb.astrouw.edu.pl/~ogle/OCVS/}{https://ogledb.astrouw.edu.pl/~ogle/OCVS/}} \citep[OCVS;][]{2014AcA....64..177S, 2019AcA....69..321S}. 

\subsection{Crossmatching and selecting RR Lyrae stars}\label{sec:VA}
\begin{figure}[ht]
\centering
\includegraphics[width=1\linewidth]{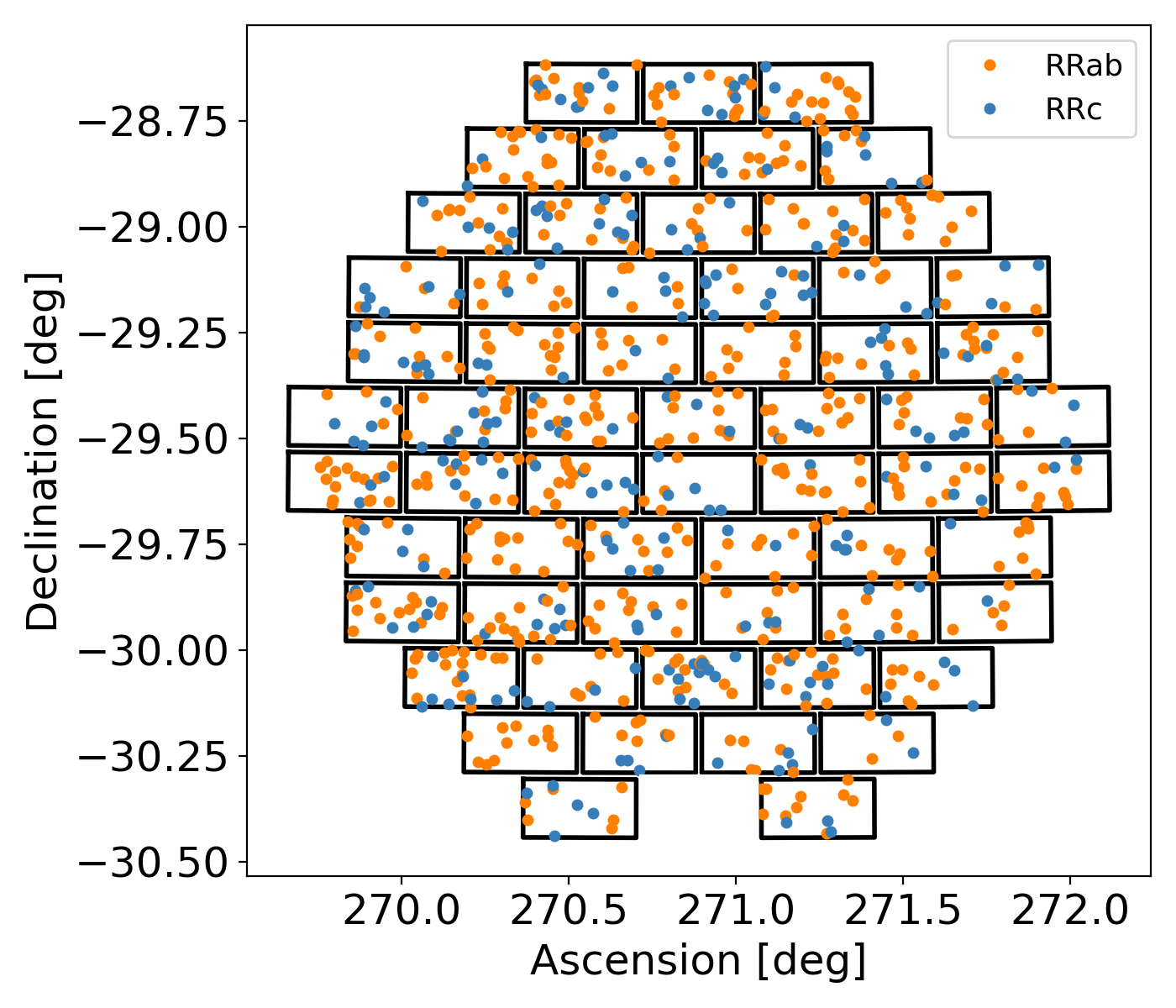}
\caption{The distribution of RR Lyrae stars in the DCP-E field was previously identified by matching our data with the OGLE Catalog of Variable Stars \citep[OCVS,][]{2014AcA....64..177S, 2019AcA....69..321S}. Orange circles represent those classified as RRab type, while blue circles represent the RRc type. The mosaic of rectangles represents the DECam footprint, where data is missing from the lower center CCD due to it being a non-functional CCD in DECam (N30).}
\label{fig:footprint}
\end{figure}

Our initial sample consisted of RRLs within 1.2 degrees of the center of the DCP-E field according to the OCVS \citep{2014AcA....64..177S, 2019AcA....69..321S}. This initial sample was further refined through a crossmatch to the catalogs produced by \texttt{photpipe} for all our images, by using the \texttt{match\_to\_catalog\_sky} method available within \texttt{astropy} \citep{2013A&A...558A..33A}. The product of this procedure is a total of 1033 RRLs (698 RRab, 331 RRc, and 5 RRd) in our data, identified and characterized according to the OCVS. Since RRd stars typically have complex light curves that are not suitable for template generation, we exclude those from further analysis. Figure~\ref{fig:footprint} depicts an approximate DECam footprint and the matched sources.

\begin{figure}[ht]
\centering
\includegraphics[width=1\linewidth]{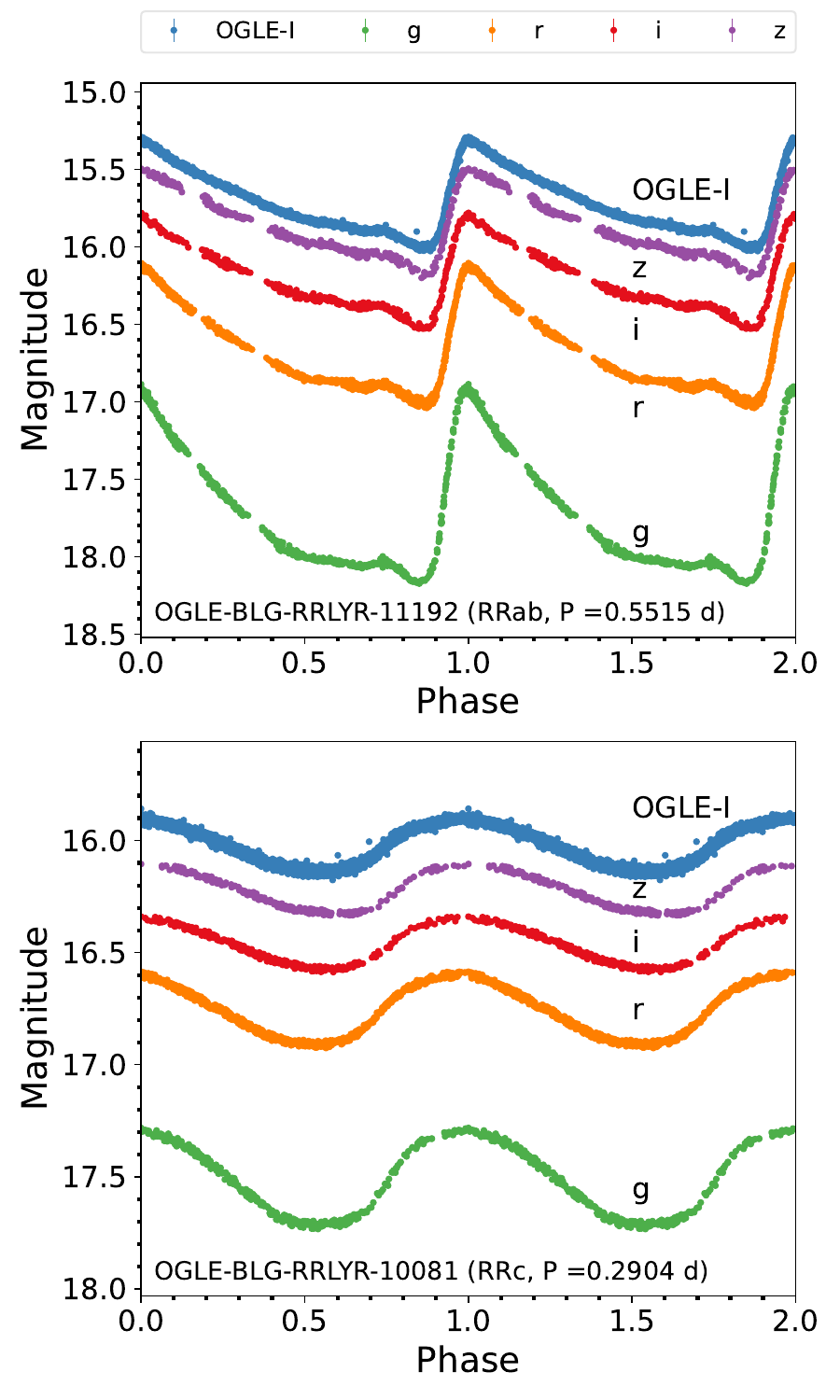}
\caption{Comparison of DECam $griz$ light curves presented in this study against the OGLE $I$-band light curve for OGLE-BLG-RRLYR-11192 (RRab) and OGLE-BLG-RRLYR-10081 (RRc). Observations in the $g$, $r$, $i$, $z$, and $I$ bands are shown in green, orange, red, purple, and blue, respectively. Note that our sigma-clipping procedure described in Section~\ref{sec:sc} has already been applied, with the exception of the OGLE $I$-band light curve.}\label{fig: ogle0}
\end{figure}

The variability of each object was independently checked in each band by visual inspection. This involved carefully examining and comparing our light curves with those from the OGLE catalog, as depicted in Figure~\ref{fig: ogle0}. By closely analyzing the patterns and characteristics of the light curves, we confirmed that most of the selected stars exhibited clear and well-defined variability. However, 20 stars in our sample were affected by nearby bright sources, which resulted in DECam light curves without apparent variability. All of these sources were discarded in the present analysis.

\begin{figure}
\centering
\includegraphics[width=1\linewidth]{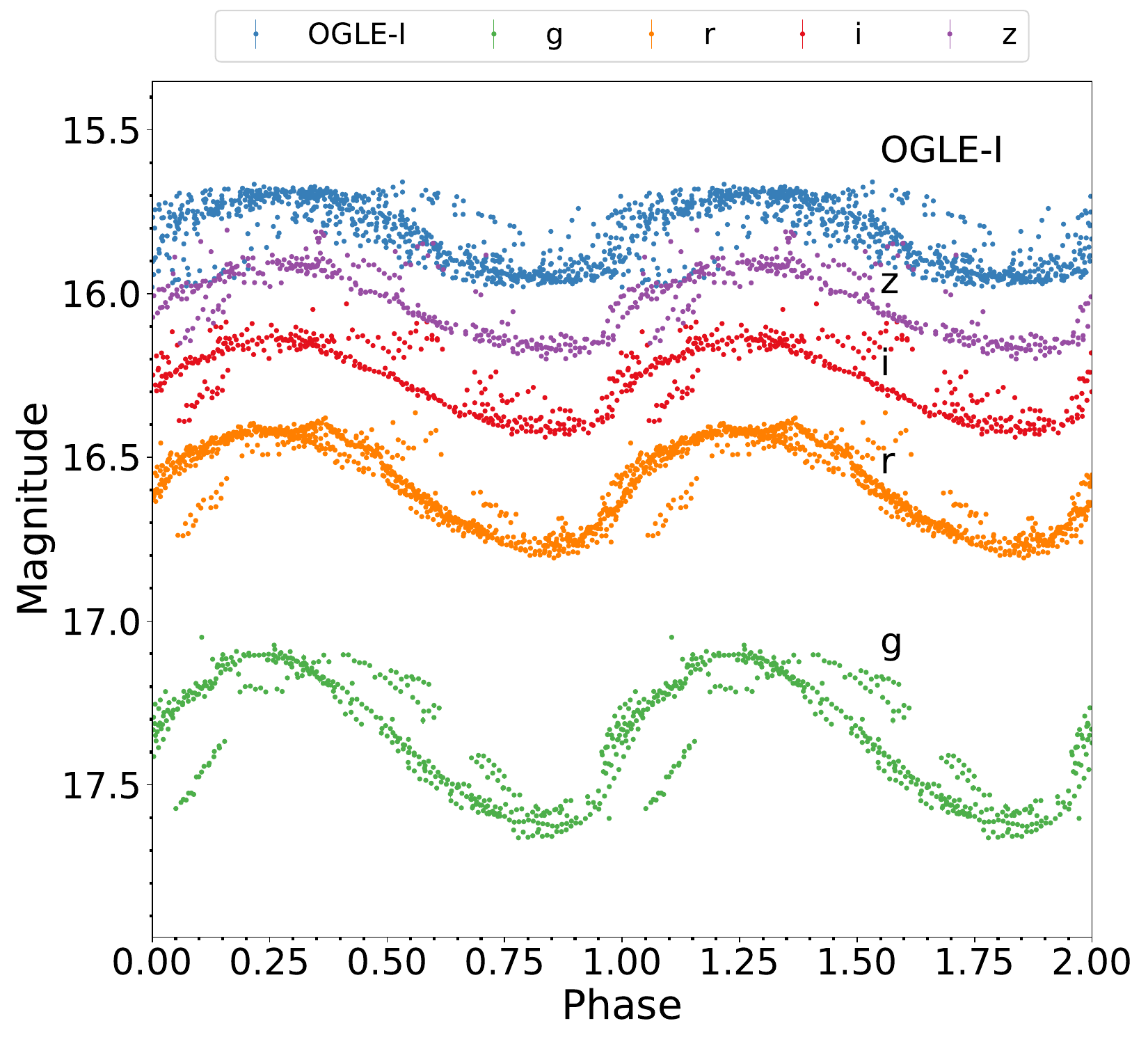}
\caption{An example of a star that shows signs of period changes, OGLE-BLG-RRLYR-13523. 
The $griz$ and OGLE $I$-band data are represented with the same color code as in Figure ~\ref{fig: ogle0}. 
Stars presenting this type of behavior were excluded from further analysis.}\label{fig: rare}
\end{figure}

We found 266 matches with the catalog from \citet{blazhko}, where 261 of these sources are classified as RRLs showing the long-term light curve modulations that are characteristic of the \citet{Blazhko1907} effect \citep[see, e.g.,][for a recent review]{Smolec2016}. The remaining sources are flagged as having noticeable period changes. These stars were set aside for future analysis. After a visual inspection, we identified an additional 270 stars exhibiting the Blazhko effect, period changes, and/or double-mode behavior, which had not been previously reported. To confirm and further investigate these findings, these stars were set aside for future analysis.
Figure~\ref{fig: rare} shows one such example. After discarding the affected stars, our final sample is comprised of 477 RRLs. 

\subsection{Sigma-clipping}\label{sec:sc}

To obtain high-quality light curve templates, outlier removal is essential.\footnote{For the present purposes, an ``outlier'' is defined as a measurement that deviates substantially from the mean (local) locus occupied by the bulk of the data for a given variable star in a time series and/or phase diagram. It should be noted that such outliers are all not necessarily spurious measurements, as certain physical processes can give rise to significant deviations from the mean light curve. In this sense, our algorithm could in principle be used in reverse, to identify outliers of potential astrophysical interest. An obvious example includes narrow eclipses in eclipsing systems containing one or more pulsating stars \citep[e.g.,][]{Soszynski2008,Soszynski2021,Debosscher2013,Udalski2015}. Other possibilities include flares, as shown, for instance, in Fig.~6 of \citet{Szabo2011} and Fig.~2 of \citet{Shi2021}, and microlensing by small bodies, as in Fig.~1 of \citet{Bond2004}. However, to the best of our knowledge, these are phenomena that have not yet been identified in the specific case of RRL stars, most likely because they are intrinsically very rare.} Unfortunately, routines such as the sigma$\_$clipping function provided by \texttt{astropy} \citep{2013A&A...558A..33A, price2018astropy} do not suit our purposes, as it is meant to remove outliers from points that are distributed normally around a mean value. In the case of light curves of variable stars, a different approach is required. For these purposes, we have devised a detrending and filtering method in which the original light curve is linearized along its length, and deviations from it are computed along the direction perpendicular to its length.  

To achieve this, a new coordinate system was introduced, wherein the light curve was parametrized based on the normalized arc length ($\ell$) and the Euclidean distance projected onto the fitted Fourier curve ($d_{\mathrm{proj}, i}$). Our Fourier fitting procedure will be described in the next section. Mathematically, the arc length of a curve represented by a function $f(x)$ can be written as

\begin{eqnarray}
l_i= \int_{x_0}^{x_i} \sqrt{\left[{(f'(x))}^2 + 1\right]} \, dx \, ,
\end{eqnarray}

\noindent where, in our case, $f(x)$ corresponds to a fitted truncated Fourier series, expressed as a function of the phase $x$. On the other hand, the Euclidean distance between an observation ($x_i$, $y_i$) and the Fourier fit is given by:

\begin{eqnarray}
d(x) = \sqrt{{(x_i - x)}^2 + {(y_i - f(x))}^2} \, .
\label{eq:distance}
\end{eqnarray}

\noindent In our case, $y_i$ corresponds to the observed magnitude at phase $x_i$. Hence, the projected Euclidean distance, i.e. the Euclidean distance perpendicular to the length of the curve, is equivalent to:

\begin{eqnarray}
d_{\mathrm{proj}, i} = \min_{0\leq x \leq 1}\{{d(x)}\} \, .
\end{eqnarray}

\noindent Thus, once we obtained $d_{\mathrm{proj}, i}$, we can compute the phase $x_{\mathrm{min}}$ at which $d_{\mathrm{proj}, i} = d(x_{\mathrm{min}})$, which in turn can be used to define the normalized arc length as follows:

\begin{eqnarray}
\ell = \frac{1}{l_{\mathrm{total}}}\int_{0}^{x_{\mathrm{min}}} \sqrt{{(f'(x))}^2 + 1} \, dx \, ,
\end{eqnarray}

\noindent where $l_{\mathrm{total}} = \int_{0}^{1} \sqrt{{(f'(x))}^2 + 1} \, dx$. 

The main idea of this new coordinate system is using $d_{\mathrm{proj}, i}$, instead of the difference between $y_i$ and $f(x_i)$, as a new criterion for the sigma-clipping procedure. In the standard sigma clipping, we have for a particular phase $x_i$ the observed magnitude $y_i$, and the prediction for the Fourier fit $f(x_i)$, and then we can exclude points where $|y_i - f(x_i)| > n\sigma$. However, this can introduce difficulties when dealing with, say, the steep rise in brightness of ab-type RR Lyrae stars, as in that part of the light curve, a small phase mismatch in the $x$ axis between observations and the Fourier fit can cause significant differences between $y_i$ and $f(x_i)$. Consequently, some points that are visually close to the Fourier fit are often improperly discarded.
Instead, by using the new coordinates $d_{\mathrm{proj}, i}$ and $\ell$, we have a more robust estimation of the apparent distance between the observed data points and predictions from the Fourier curve.\footnote{In the case of high-amplitude variables, the user may choose to normalize the magnitude values used in eq.~\ref{eq:distance}, to provide a more ``impartial'' assessment of distances in this new coordinate system.} Essentially, for a certain normalized arc length $\ell_i$, we have the projected distance of the observation to the Fourier curve $d_{\mathrm{proj}, i}$. We can exclude those points that satisfy $d_{\mathrm{proj}, i} > n\sigma$, where now $\sigma$ is the standard deviation of the distribution of distances, computed perpendicularly from the linearized light curve.
It is important to note that our method is iterative. Points near the rising branch that appear offset by several sigma in magnitudes from the Fourier fit may still be valid data points, as the Fourier fit itself is not perfect. Our method accounts for the imperfections in the Fourier fit during each iteration, ensuring that points a few sigma away in magnitude from the current curve are not necessarily removed. This approach provides a more accurate and reliable selection of data points. 

To estimate the error in the new coordinates, we calculated the angle $\theta_i$, given by

\begin{eqnarray}
\theta_i =\arctan(dm_i/dx) - \frac{\pi}{2} \, ,
\end{eqnarray}

\noindent where $dm_i/dx$ corresponds to the magnitude derivative obtained on the basis of the Fourier fit at each point in phase. This angle is used to calculate $e_{l,i}$ and $e_{d,i}$, which represent the projected errors in our new coordinates, as follows: 

\begin{eqnarray}
e_{l,i}  &=& e \cdot \cos(\theta) \, , \\
e_{d,i} &=& e \cdot \sin(\theta)\, ,
\end{eqnarray}
\noindent where $e$ is the error in the original coordinate system,  i.e., the error in the photometry, as we assume the periods are known and the times are essentially error-free. 

\begin{figure*}[ht!]
\centering
\includegraphics[width=0.95\linewidth]{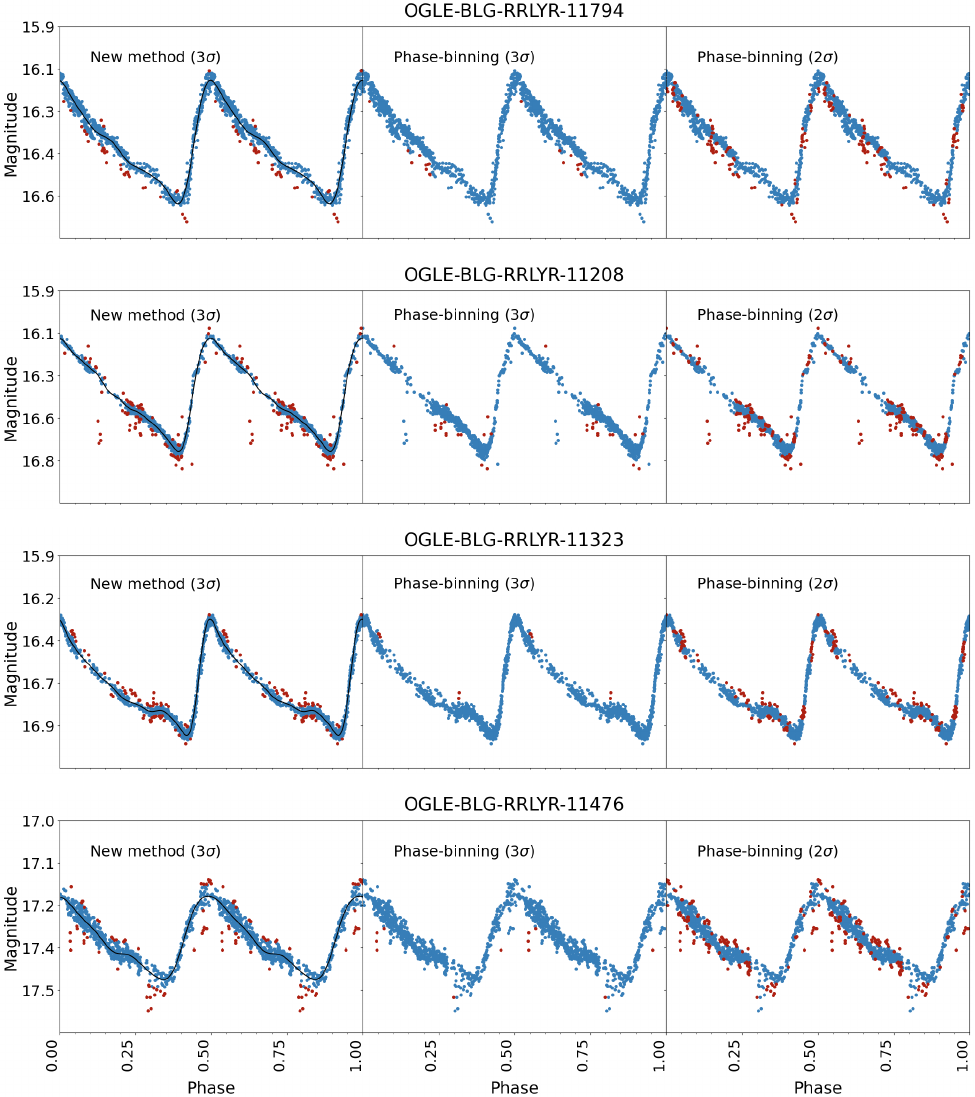}
\caption{Example of the sigma-clipping technique proposed in this work for six RRab stars, where the black line depicts the Fourier fit in the $r$ band. The yellow dots represent the remaining data points, and the red dots represent the data points removed using the $3\sigma$ threshold with five iterations (left panel). The central and right panel shows the phase-binning method using bins of 0.05 width (in phase) and a $3\sigma$ and $2\sigma$ threshold with five iterations, respectively.  
}\label{fig:1}
\end{figure*}

To evaluate the performance of our method, in Figure~\ref{fig:1} we plot the light curves in the $r$-band for six RRab stars. Specifically, we compare the outlier removal achieved with our method with the one obtained using phase binning. The latter is another common tool that performs outlier removal by binning the light curve in phase and analyzing the distribution of magnitudes in each bin. The sigma-clipping method proposed here was applied using a $3\sigma$ rejection threshold and 5 iterations. These parameters were selected after testing with $1, 2$, and $3\sigma$ rejection thresholds, and performing between 1 and 6 iterations. It was found that, with $1\sigma$ or $2\sigma$ thresholds, the method was overly strict, removing data points that by visual inspection did not appear to be outliers. Conversely, the $3\sigma$ option effectively identified and removed outliers with greater accuracy. Additionally, increasing the number of iterations beyond 5 did not affect the final result, leading to the selection of this value. It should be noted that the choice of these parameters depends on the characteristics of the data being analyzed.
On the other hand, it is important to consider that, when using the phase-binning sigma-clipping method, if we apply a $3\sigma$ threshold, generally not all outliers are removed, as can be seen in the middle column of Figure~\ref{fig:1}. For this reason, $2\sigma$ is often used. However, when using $2\sigma$, the generic sigma-clipping method tends to remove data that are not outliers, as shown in the third column. This becomes even more significant when applying a Fourier fit, as these outliers can significantly affect the behavior of the fit. Therefore, our routine performed better job at removing outliers than does the phase-binning method.  

\section{The $griz$ RR Lyrae templates}\label{sec:3}

Our multi-band light curve templates are provided in the form of truncated Fourier series. Our approach to obtain the corresponding coefficients is described in this section. 

\subsection{Fourier decomposition technique}\label{sec:fourier}
As with any regular mathematical function, light curves can be expressed as a sum of cosine and sine series \citep[e.g.][]{1986A&A...170...59P,2009A&A...507.1729D,2013arXiv1309.4297N}, as follows:  

\begin{eqnarray}
m(t) &=& A_{0}+\sum_{i=1}^{N} a_{i} \cos \left(i \omega\left(t-t_{0}\right)\right)  \nonumber \\
& +& \sum_{i=1}^{N} b_{i} \sin \left(i \omega\left(t-t_{0}\right)\right) \, , 
\end{eqnarray}

\noindent where $m(t)$ represents the observed magnitude at time $t$, $A_{0}$ is the mean magnitude, $N$ is the order of the fit, $a_{i}$ and $b_{i}$ are the amplitude components of the $i^{\text{th}}$ harmonic, $\omega = 2 \pi / P$ is the angular frequency, $t_{0}$ is the epoch of maximum light, and $P$ is the period of the star in days, as adopted from OCVS \citep{2014AcA....64..177S, 2019AcA....69..321S}. Secondly, we can fold the time observation into phase as

\begin{eqnarray}
\Phi = \frac{\left(t-t_{0}\right)}{P} -
\left \lfloor \frac{t-t_{0}}{P} \right \rfloor .
\end{eqnarray}

\noindent By employing the properties of trigonometric functions, $m(t)$ can also be represented as follows: 

\begin{eqnarray}
m(t) = A_{0} + \sum_{i=1}^{N} A_{i} \cos (2 \pi i \Phi(t) + \phi_i) \, , 
\end{eqnarray}

\noindent where $i>1$, $A_i = \sqrt{a_i^2 + b_i^2}$, and $\tan(\phi_i) = - (b_i / a_i)$. The {\em relative} Fourier parameters \citep{Simon1981}  are defined as 

\begin{eqnarray}
R_{i 1} &=& \frac{A_{i}}{A_{1}} , \\
\phi_{i 1} &=& \phi_{i}-i \phi_{1} \, . 
\end{eqnarray}
 
\noindent Both these parameters can also be used to describe the light curve shape \citep[e.g.,][]{2009A&A...507.1729D}. 

Determining the optimal number of terms in the Fourier decomposition of an individual light curve is complex. It is a crucial step because, if the value of $N$ is too small, it may not capture all the essential features of the light curve; setting it too large, on the other hand, can result in overfitting, where the noise is modeled \citep{1986A&A...170...59P,2009A&A...507.1729D}. In this work, we initially determined the $N$ value using Baart's criterion \citep{10.1093/imanum/2.2.241,1986A&A...170...59P}, and then adjusted the resulting $N$ value through visual inspection of each light curve. It is important to note that Baart's criterion does not account for errors in the photometry, which is one reason why adjusting the $N$ value is necessary. Table~\ref{tab:n} displays the median $N$ values for each band, specifically for the RRab and RRc stars. RRc stars have smaller values of $N$ since their light curves exhibit more sinusoidal behavior than do the light curves of RRab stars. As a result, the morphology of the light curves of RRc stars is generally less complex. In like vein, the $N$ value decreases with increasing effective wavelength, which again reflects the fact that light curves become increasingly more sinusoidal as one approaches the infrared regime from the visual \citep[e.g.,][]{PS}. 

\subsection{Sample selection}
To derive a precise $griz$ light-curve templates, we selected variables from the sample of 477 RRLs that satisfied the following criteria:

\begin{table}
\centering
\caption{Median $N$ values obtained in the Fourier fits to the $griz$ light curves of RRab  and RRc stars.}
\label{tab:n}
\begin{tabular}{lccccc}
\hline\hline
Type & \multicolumn{4}{c}{$N$}  \\ 
\cline{2-5}
     & $g$ & $r$ & $i$ & $z$  \\ 
\hline 
RRab  & 9 & 8 & 6 & 6  \\ 
RRc& 3 & 3 & 2 & 2  \\ 
\hline
\end{tabular}
\end{table}

\begin{enumerate}
\item {At least fifteen observations in each band, as in \citet{sesar}, to help ensure good phase coverage. In our final sample, the minimum number of observations per band is 48 for the RRab and 30 for the RRc. The mean number of observations per star is 341, the median is 291, and the standard deviation is 137 for the RRab stars, demonstrating robust observational coverage across the sample. For the RRc stars, the mean number of observations per star is 322, the median is 288, and the standard deviation is 148, confirming this group's excellent phase coverage.}

\item No irregularities in the Fourier fit caused by poor phase coverage. We removed RRL templates affected by such irregular behavior by visual inspection. 
\end{enumerate}

\begin{figure}
\centering
\includegraphics[width=1\linewidth]{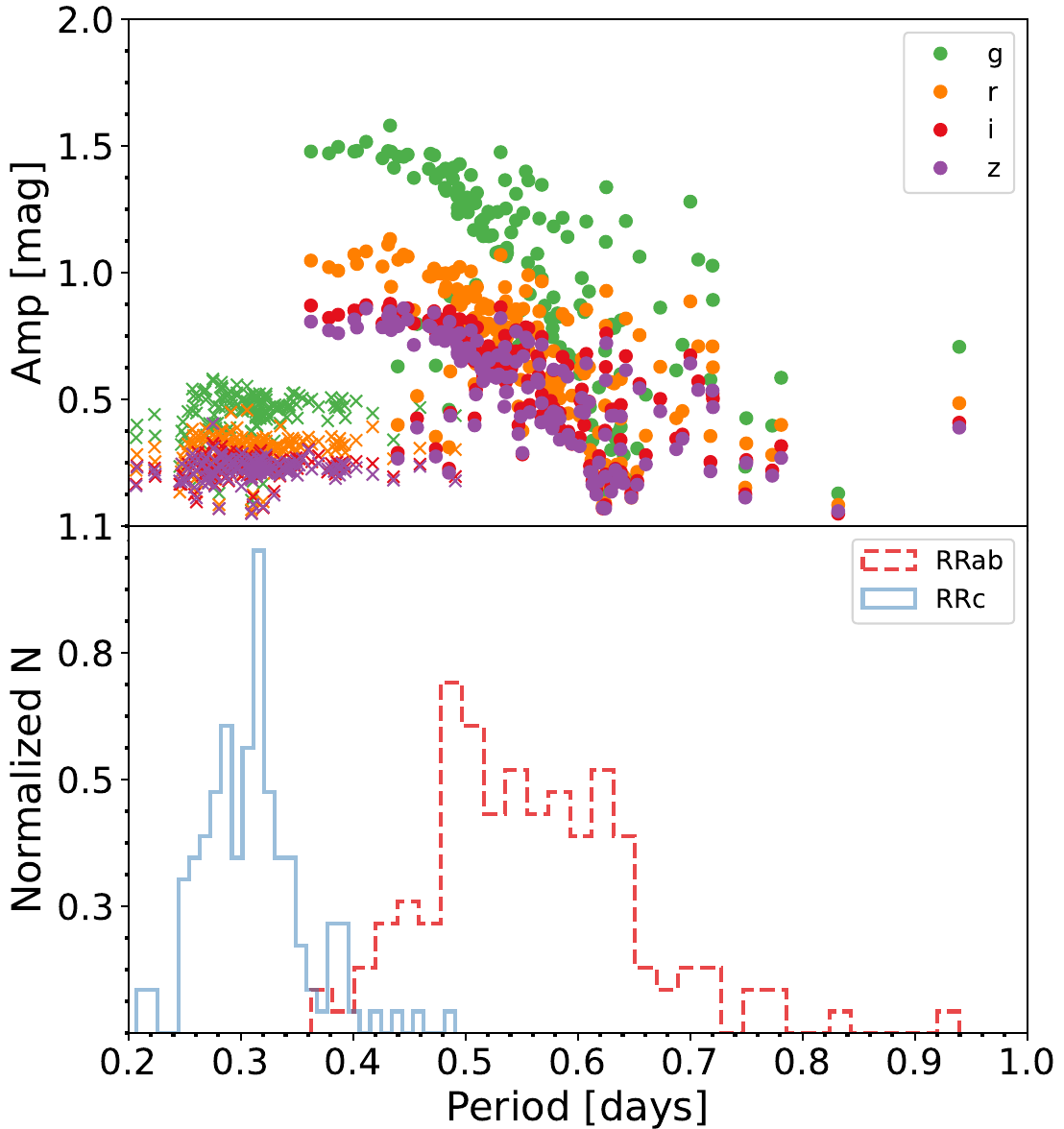}
\caption{Period-amplitude diagram in the $griz$ bands (top panel) and period distribution (bottom panel). Circles and crosses represent RRab and RRc stars, respectively, color-coded by bandpass according to the color scheme shown in the upper-right corner of the top panel.}\label{fig:periods}
\end{figure}

\begin{figure*}
\centering
\includegraphics[width=1\linewidth]{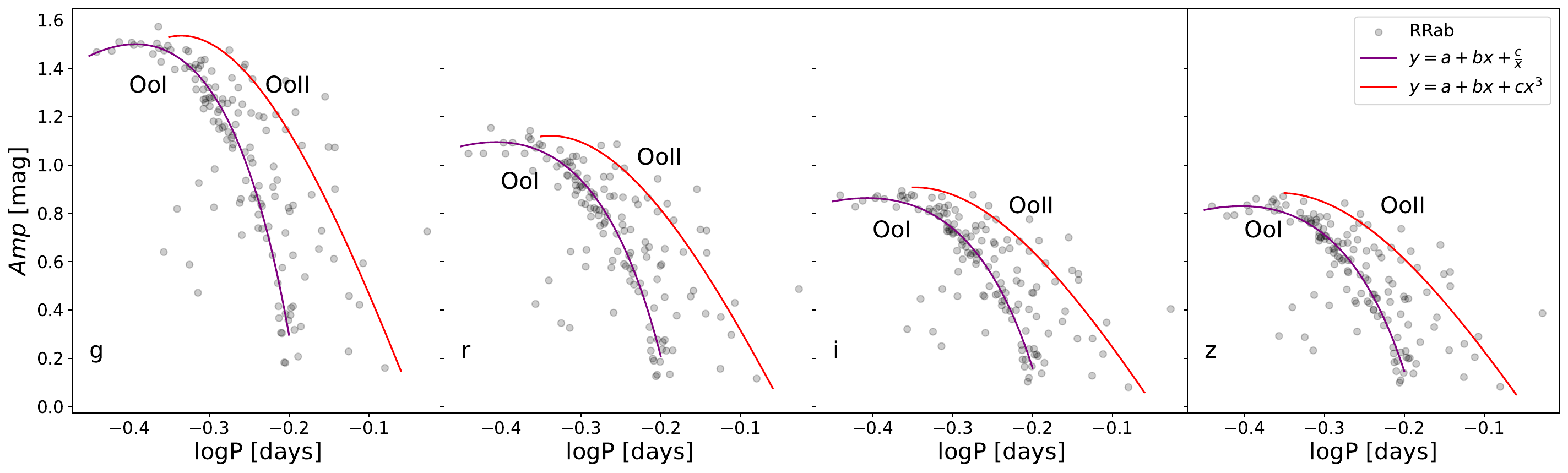}
\caption{Period-amplitude diagram for the $griz$ bands of RRab stars. The purple and red solid lines represent the new analytical relations for the OoI and OoII loci, as provided in equations~\ref{eq:OoI} and \ref{eq:OoII}, respectively.}\label{fig:Oogroup}
\end{figure*}

Upon applying these criteria, our sample is reduced to 280 RRL, 136 of which correspond to RRab and 144 to RRc variables. Note that four templates are available for each star, i.e., one per each available $griz$ band. The period-amplitude (or Bailey) diagram and the period distributions for these RRL stars are shown in Figure~\ref{fig:periods}. To define a polynomial relation that adequately describes the loci of Oosterhoff type I (OoI) and type II (OoII) stars \citep[see, e.g.,][for the definition of Oo types]{PS}, we followed the procedure described in \citet{2019Prudil}. First, we binned our stellar sample based on their amplitudes, computed kernel density estimates as a function of the period for each amplitude bin, and determined the maximum in each bin. We then fit these maxima with a curve that best meets the expected physical conditions and describes the average behavior of the data. Second, we applied the conditions prescribed by \citet{2008Miceli} to distinguish between the two Oosterhoff groups, using a fixed threshold of 0.045 in logarithmic period shift, calculated at fixed $g$-band amplitudes. Next, we explored suitable analytical representations of these data, finding that the following expressions conveniently describe the average loci occupied by the OoI and OoII components in all four bands:

\begin{equation}
\log P = a + b \, {\rm Amp} + \frac{c}{\rm Amp}, \,\,\,\,\, {\rm (OoI)} 
\label{eq:OoI}
\end{equation}
\begin{equation}
\log P = a + b \, {\rm Amp} + c \, {\rm Amp}^{3}; \,\,\,\,\, {\rm (OoII)} \\ 
\label{eq:OoII}
\end{equation}

\noindent their respective $a$, $b$, $c$ coefficients are shown in Table~\ref{table:Oo}. The results of this procedure are shown in Figure~\ref{fig:Oogroup}.

\begin{table}[h!]

\caption{Coefficients for the analytical relationships describing the Oosterhoff loci in the $griz$ bands.}
\label{table:Oo}
\small
\centering
\begin{tabular}{cccc|cccc}
\toprule
\hline
\multicolumn{4}{c|}{Oo I} & 
\multicolumn{4}{c}{Oo II} \\
\cmidrule(r){1-4} \cmidrule(l){5-8}
Band & $a$ & $b$ & $c$ & Band & $a$ & $b$ & $c$ \\
\midrule
$g$ & 6.61 & 6.52 & 1.00 & $g$ & -0.36 & -8.47 & 25.17 \\
$r$ & 4.49 & 4.18 & 0.69 & $r$ & -0.30 & -6.30 & 18.40 \\
$i$ & 3.54 & 3.29 & 0.54 & $i$ & -0.23 & -4.93 & 13.63 \\
$z$ & 3.48 & 3.28 & 0.54 & $z$ & -0.23 & -4.70 & 12.36 \\
\bottomrule
\end{tabular}
\tablefoot{The four columns on the right show the coefficients for the OoII locus (eq.~\ref{eq:OoII}), whereas the four on the left contain the coefficients for the OoI locus (eq.~\ref{eq:OoI}).}
\end{table}

\subsection{The griz RR Lyrae templates}\label{sec:templs}
\begin{figure*}
\centering
\includegraphics[width=1\linewidth]{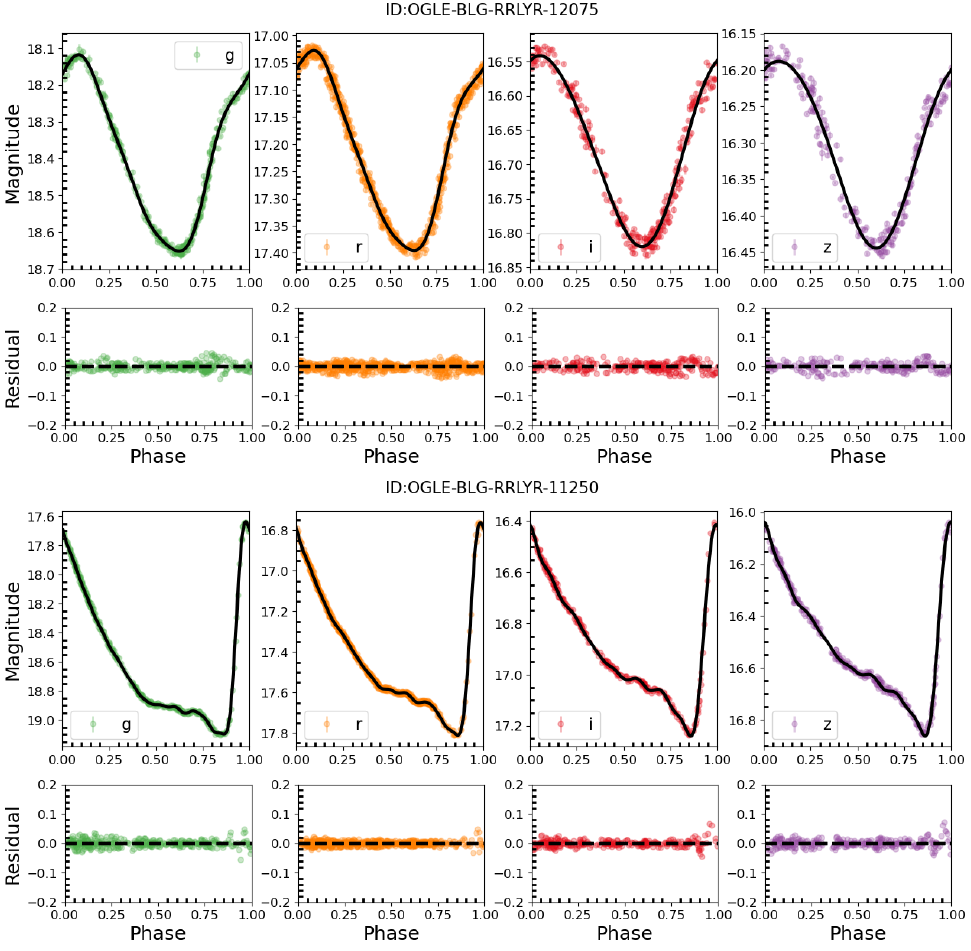}
\caption{Examples of multiband templates for OGLE-BLG-RRLYR-12075, an RRc (two upper panels), and OGLE-BLG-RRLYR-11250, an RRab (two bottom panels). 
In the main light curve plots, the points represent individual measurements in, from left to right, the $griz$ bands. 
Underneath each set of light curves, the residuals between the derived template (Fourier fit) and the observed magnitude data for both RRL subclasses are shown.}\label{fig:residuos}
\end{figure*}

The above procedure was applied to our full sample of RR Lyrae stars, and templates derived accordingly; examples are shown in Figure~\ref{fig:residuos}, where the derived multi-band templates are overplotted on the empirical data for the RRab star OGLE-BLG-RRLYR-11250 (upper panels) and the RRc star OGLE-BLG-RRLYR-12075 (bottom panels). 
This figure reveals excellent agreement between our modeling and the data. As expected, the largest residuals are still found on the rising branch, due to the inherently rapid light curve evolution in this phase of the pulsation cycle. However, this does not affect the overall quality of the fits.

Careful comparison between our templates and the data they are modeled after reveals a tendency for increased fluctuations for stars with fewer data points, smaller amplitudes, and/or fainter magnitudes. These trends are as expected, and explain the slightly larger residuals in the case of the redder, smaller-amplitude passbands shown in Figure~\ref{fig:residuos}. This increased scatter notwithstanding, the corresponding templates still properly describe the expected light curve behavior displayed by both RRab and RRc stars. 

\begin{figure*}
\centering
\includegraphics[width=1\linewidth]{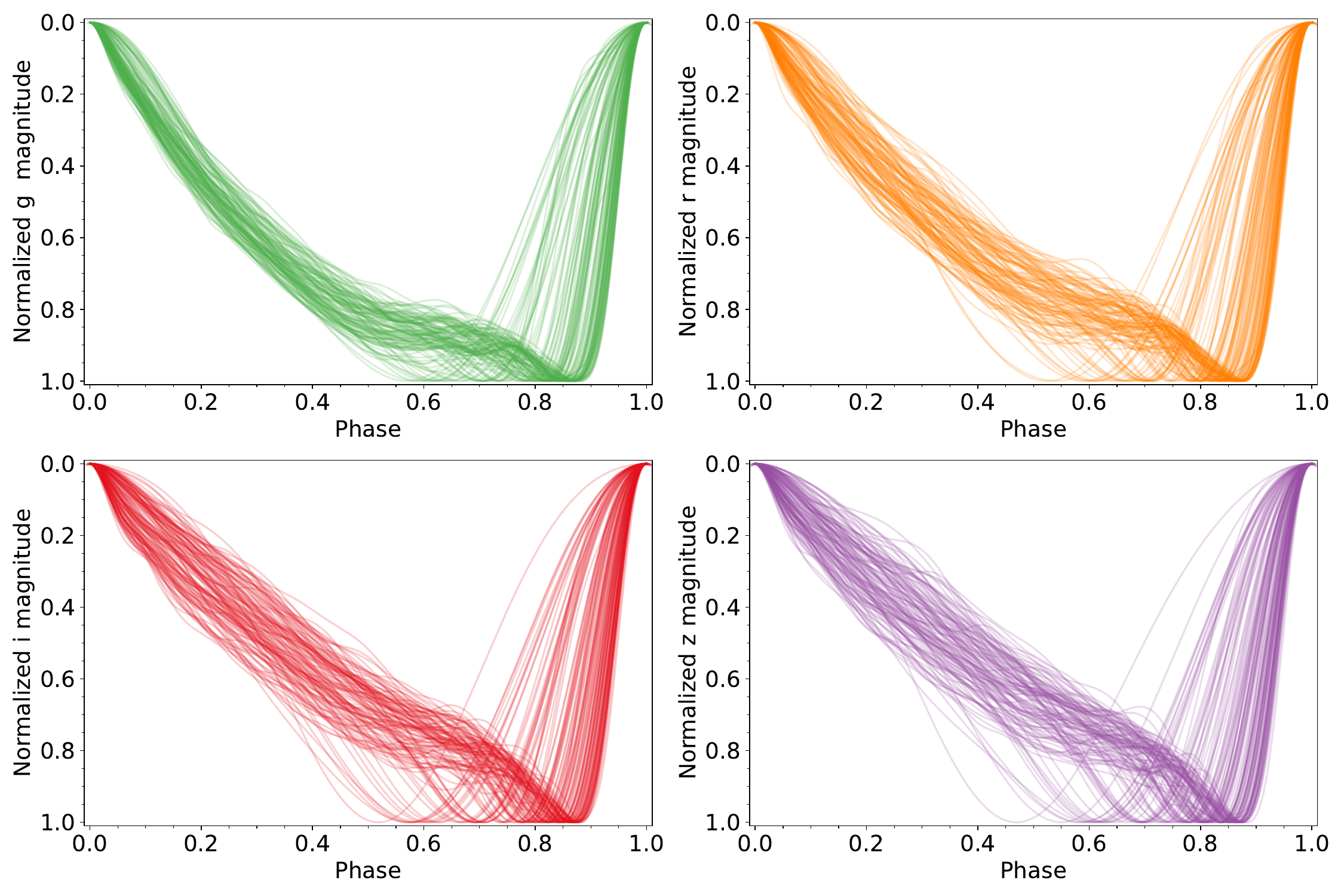}
\caption{Phase against normalized magnitude for our 136 RRab templates. From left to right: the top panels contain $g$ and $r$ bands, while the bottom panels show $i$ and $z$. The bands are shown in green, orange, red, and purple,
respectively.}\label{fig: t0}
\end{figure*}

\begin{figure*}
\centering
\includegraphics[width=1\linewidth]{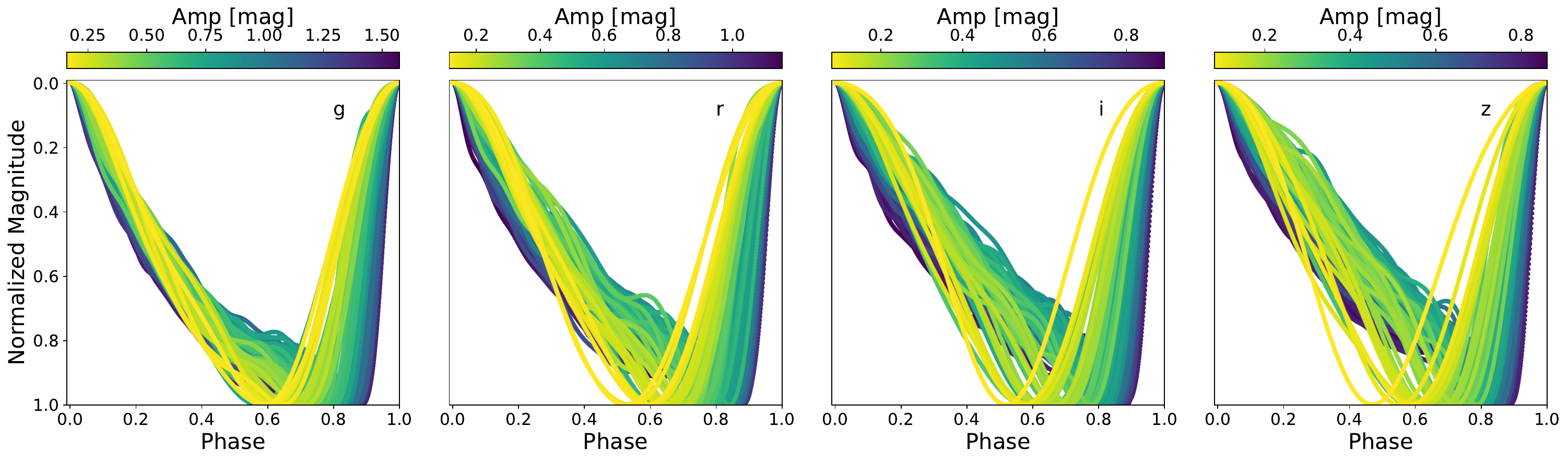}
\caption{Templates obtained for RRab stars in the $griz$ bands (from left to right), color-coded according to their real amplitudes following the color bar on the top of each panel.}\label{fig:amp}
\end{figure*}

Figures~\ref {fig: t0} and \ref{fig: t1} show our final set of templates per band. 
In these plots, the magnitude range has been rescaled to the range $[0, 1]$, so that the amplitudes are normalized to one. 
As shown in Figure~\ref{fig:periods}, the templates for RRab stars cover a period range from 0.36 to 0.93 days, while the templates for RRc stars span from 0.20 to 0.49 days. 
Figure~\ref{fig:amp} shows, in addition, a notable trend where the minimum of the light curve is reached at later phases (implying smaller rise times) for greater amplitudes, as expected for RRab stars. 

We provide two quality indicators to facilitate the use of these templates. First, we visually classify the templates according to their quality ($Q$). A template with a quality rating $Q = 2$ has magnitude values well sampled across all phases, and the template follows the observational trends. Then, a quality rating $Q = 1$ suggests that, although 
phase coverage may be relatively limited, any phase gaps do not adversely affect the final template morphology. Finally, the second indicator ($Q^\prime$) evaluates the quality of phase data by dividing the phase into a series of equal-sized bins, covering the entire range from 0 to 1 in steps of 0.01, which results in 100 bins. These bins are evenly distributed across the phase space. Within each bin, the function calculates residuals, which are the differences between the observed and interpolated magnitudes, normalized by the photometric errors. For each bin, it computes the standard deviation of these residuals, reflecting the spread of the errors. The spread is then used in an inverse logistic function to derive a quality score for that bin. Finally, the overall quality indicator is calculated as the average of the quality scores across all bins, providing a comprehensive measure of the consistency and accuracy of the phase data. Typical quality values for the best RRL templates are usually around 0.8 to 1. On the other hand, lower quality templates generally have quality values below 0.5.

The templates and quality indicators are available via Zenodo\footnote{\href{https://doi.org/10.5281/zenodo.11261281}{https://doi.org/10.5281/zenodo.11261281}} 
and Github\footnote{\href{https://github.com/KarinaBaezaV/Multiband-templates}{https://github.com/KarinaBaezaV/Multiband-templates}} repositories.

\subsection{Comparison against Sesar et al. (2010)}
\label{sec:sesar}
To test the consistency and reliability of our templates, we compare against the ones previously presented by \citet{sesar}. Unlike ours, the \citet{sesar} templates were built for RRLs found in the Stripe 82 region of the SDSS \citep{2015ApJS..219...12A, York_2000}. They provide a set of 
12, 23, 22, 22, and 19 templates in the $ugriz$ bands, respectively.

In contrast, they offer only 1 and 2 templates for RRc stars in each of the $uz$ and $gri$ bands, respectively.

\begin{figure*}
\centering
\includegraphics[width=1\linewidth]{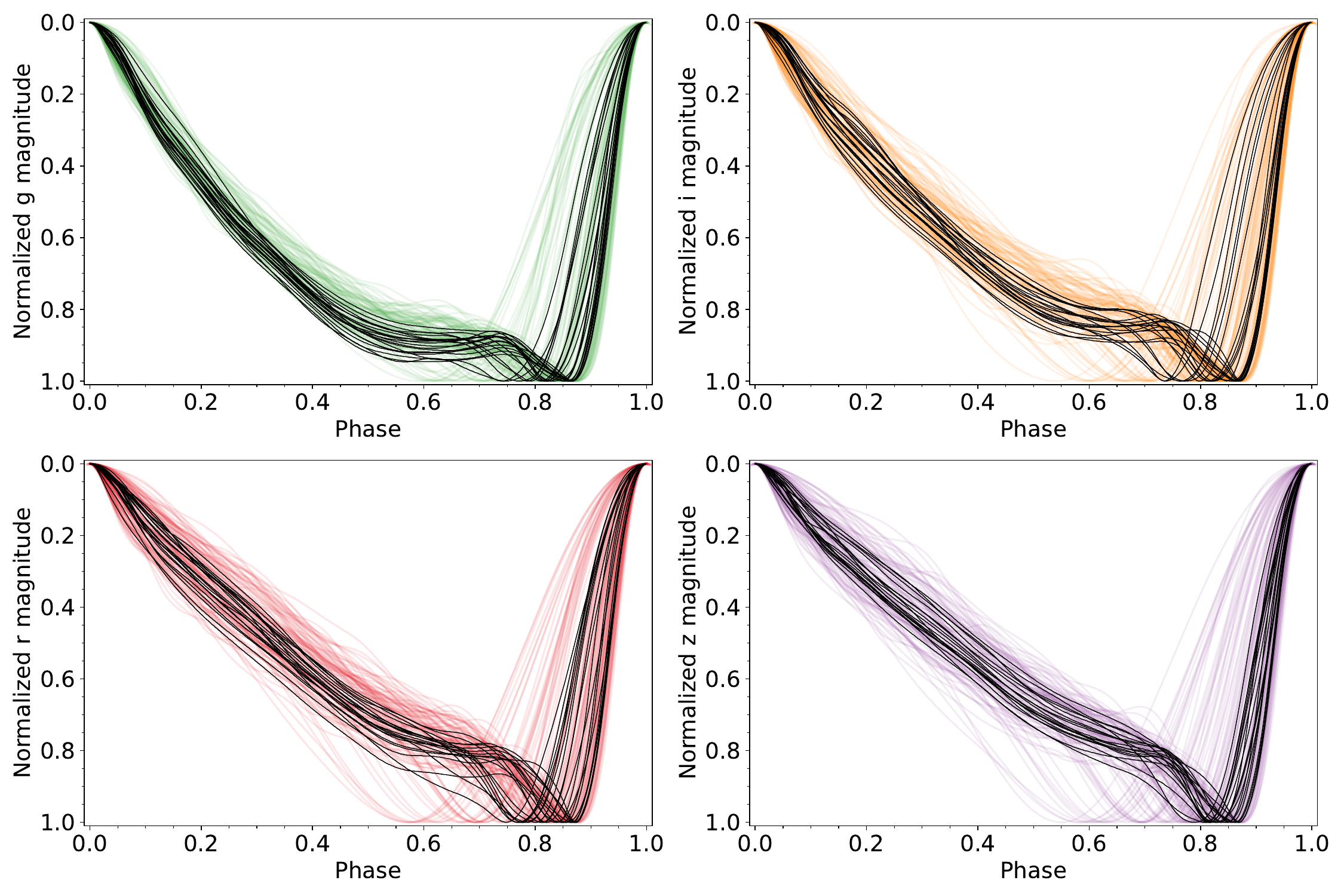}
\caption{Comparison of the templates obtained in this study for the $griz$ bands and those from the study by \citet{sesar} (black) in the SDSS system. Note that all template shapes presented in \citet{sesar} are contained as subsets of our templates.}\label{fig: sesar2}
\end{figure*}

Figure~\ref{fig: sesar2} shows our templates obtained for the $griz$ bands, and the templates from the \citet{sesar} study shown in black. To carry out this comparison, our DECam magnitudes were transformed to the SDSS system\citep{1996AJ....111.1748F} by means of the following procedure. 
 
First, we apply photometric transformations from the DECam system to Pan-STARRS \citep{2016arXiv161205560C}, using the coefficients given in \citet{2018ApJS..234...39S}. Next, we perform the transformation from Pan-STARRS to Sloan, following the procedure described in \citet{2012ApJ...750...99T}. 

Comparing the overall trends, it is evident that our templates closely follow the general shape and behavior of the \citet{sesar} templates in each band. The rising and descending branches and the characteristic humps and bumps are largely consistent between the templates, 
but ours capture more nuanced features of RRab light curves. These include, for instance, a bump at a phase around 0.65, also seen in the light curves shown in \citet{2017ApJ...834..160N}, which are not as clearly represented in the \citet{sesar} template set. 
 
Indeed, our template sample provides a more comprehensive representation compared to their templates as it encompasses a broader range of amplitudes, periods, rise times,including also RRab stars with minima at phases below 0.70. Another critical distinction pertains to the methodology adopted for constructing the templates; in \citet{sesar}, prototype templates were smoothed by averaging over similarly-shaped light curves a necessity in their case, as light curves from the SDSS Stripe 82 were used, whose quality are not comparable to those used in our analysis, in terms of phase coverage, signal-to-noise ratio, and total number of data points per light curve alike.

Accordingly, in our work we opted against averaging the templates. The outstanding quality of our individual light curves underpins this choice. Averaging templates, especially those derived from such high-quality data, risks obfuscating unique features or subtle characteristics of individual curves. By eschewing averaging, we retain the details and variations inherent to each band, thus ensuring templates that more closely follow the observed variety of light curve shapes.

\section{Fourier parameters}\label{sec:4}
Fourier parameters play a crucial role in analyzing and characterizing periodic signals. One of their key advantages is their ability to quantify the shape and variability of periodic signals \citep{1971A&AS....4..265S,1996A&A...312..111J}. Furthermore, in the case of RR Lyrae stars, Fourier parameters can help differentiate between the RRab and RRc subtypes \citep[see, e.g.,][]{1985ApJ...299..723S,2003A&A...398..213M,sesar,Mullen2021,Mullen2022,nonblazhko}.

\begin{figure*}
\centering
\includegraphics[width=1\linewidth]{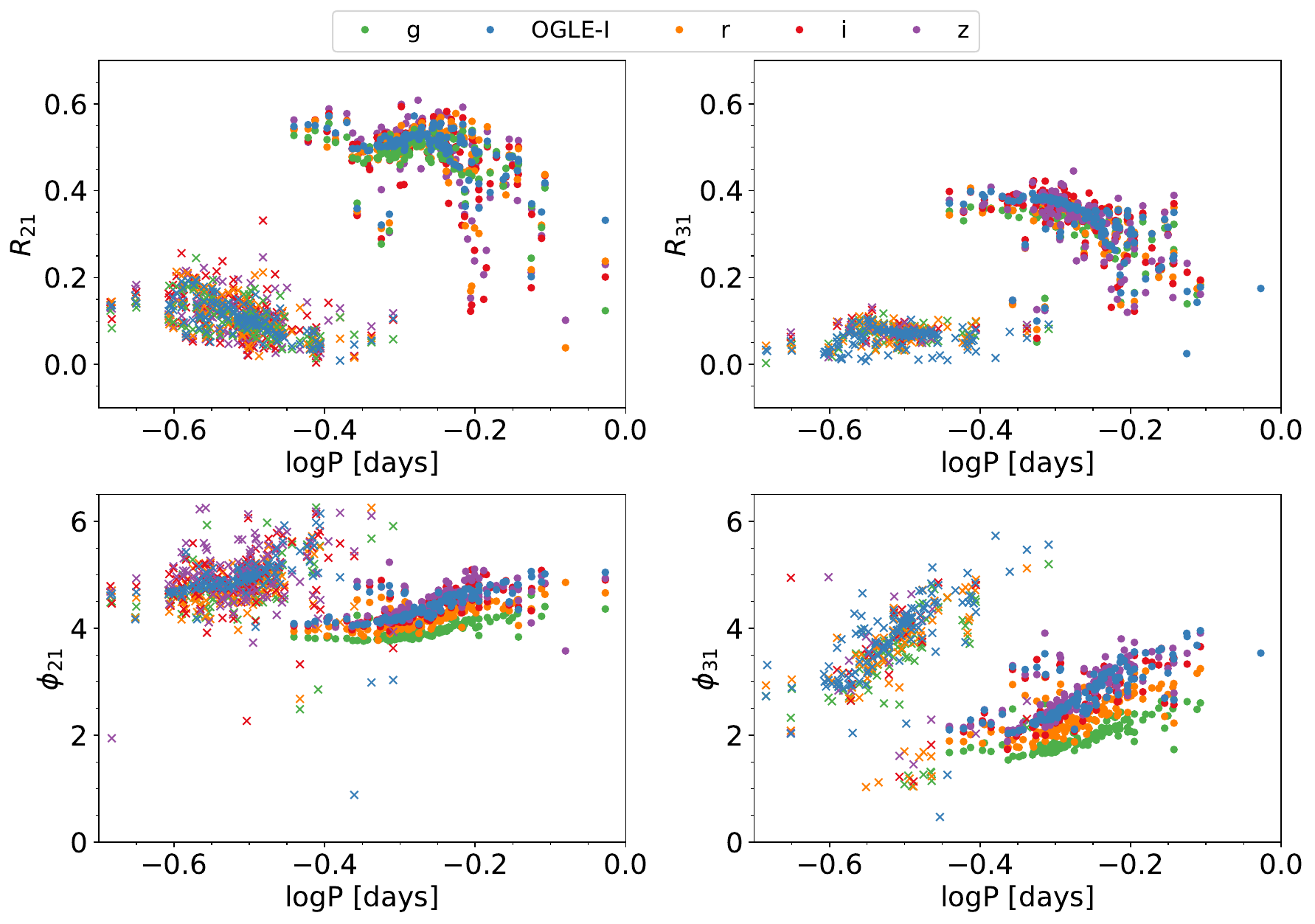}
\caption{Fourier parameters $R_{21}$ (upper left), $\phi_{21}$ (bottom left), $R_{31}$ (upper right), and $\phi_{31}$(bottom right) as a function of the logarithm of the period, as obtained in this study for the $griz$ bands, along with those reported by OCVS  in the $I$ band for RRab (circles) and RRc (crosses) stars. Our $griz$ data and OGLE's $I$-band data are represented using the same color scheme as in Figure~\ref{fig: ogle0}.}\label{fig: griz12}
\end{figure*}

The light curve parameters of our sample of RRLs obtained from the best-fit $griz$ templates are listed in Table~\ref{table:properties}. Figure~\ref{fig: griz12} depicts four subplots for the $griz$ bands, illustrating the relationship between the low-order Fourier parameters ($R_{21}$, $\phi_{21}$, $R_{31}$, and $\phi_{31}$) and the logarithm of the period, and compares them with those reported by OCVS for RRab and RRc stars. Firstly, for RRab stars, it is evident that $R_{21}$ and $R_{31}$ generally decrease as the period increases. Moreover, we observe an increase in $\phi_{21}$ and $\phi_{31}$ with longer periods. To analyze these trends in a more quantitative way, in Figure~\ref{fig:PearsonRRab} we analyze the interrelation among all light curve parameters computed in this paper, quantified by means of their respective Pearson correlation coefficients and displayed in the form of a ``confusion matrix''. As can be clearly seen in this figure, both parameters $R_{21}$ and $R_{31}$ show negative correlations with the period across all bands. 
 
On the other hand, for $\phi_{21}$ and $\phi_{31}$, the correlation coefficients are all positive with the period.
These findings are in agreement with similar results from previous studies, 
specifically using SDSS filters  \citep{2017ApJ...834..160N}. Other studies reported comparable trends, although they employed different bandpasses \citep{2009A&A...507.1729D,humps, nonblazhko}.

Our distribution of $R_{21}$ and $R_{31}$ values is largely consistent with that reported by \citet{nonblazhko} for non-Blazhko stars. In addition, we observe a trend in that $\phi_{21}$ and $\phi_{31}$ are higher in redder bands, similar to \citet{2017ApJ...834..160N}. This behavior is expected, considering the fact that the amplitudes decrease and the light curves become more sinusoidal as one moves toward redder bands such as $iz$. 
Secondly, for RRc stars, we observe a similar behavior;  
however, the correlation coefficients for RRc stars are much smaller 
compared to those for RRab stars. Comparing these results with \citet{2017ApJ...834..160N}, our distribution appears sharper and clearer as the period increases. Additionally, when comparing the values obtained in this work with those reported by OCVS, we observe a high degree of similarity. In particular, the results for the $i$ band in this study are similar to those reported by OCVS in the $I$ band. Figure~\ref{fig:PearsonRRc} shows the Pearson correlation coefficients obtained for the RRc stars, which support the above discussion.

\begin{figure*}[ht]
\centering
\includegraphics[width=1\linewidth]{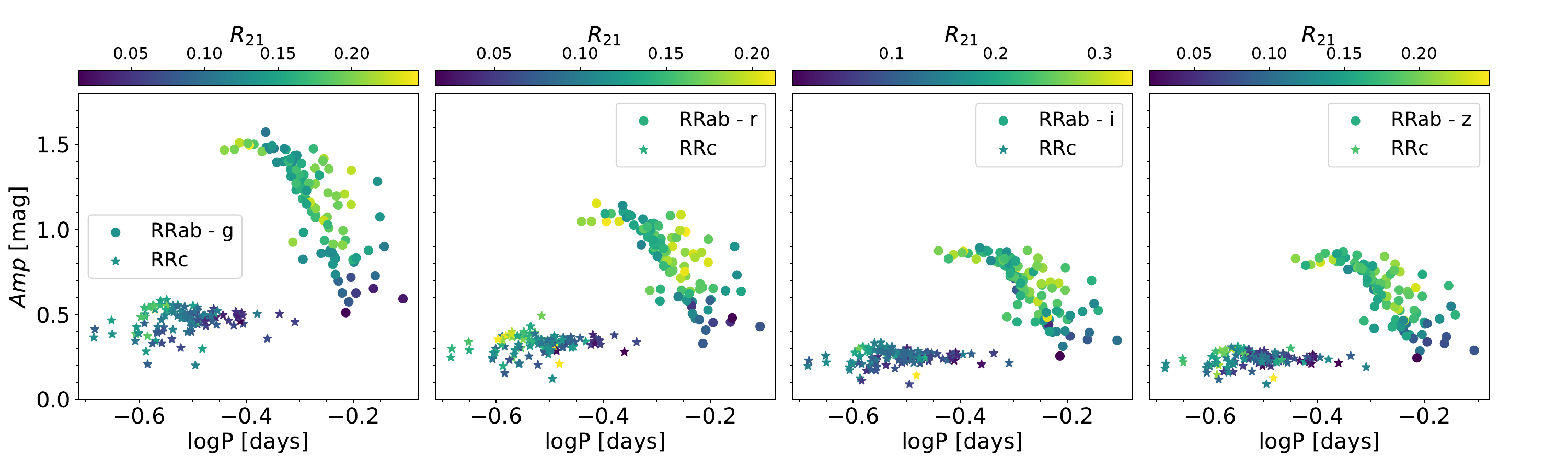}
\caption{Total amplitude as a function of $\log(P)$ in (from left to right) the $griz$ bands for RRab and RRc stars, represented by circles and stars, respectively. The color of the symbols indicates the value of the Fourier parameter $R_{21}$, as shown by the color bar at the top.}\label{fig: AR}
\end{figure*}

Additionally, we observe a noticeable correlation between the amplitude and the logarithm of the period, dependent on $R_{21}$ in the four bands. As the period increases, the value of $R_{21}$ decreases on average, accompanied by a decrease in the amplitude. This suggests that RRLs with longer periods and smaller amplitudes tend to show lower values of $R_{21}$ in the $griz$ bands, as illustrated in Figure~\ref{fig: AR} and supported by other studies \citep{2009A&A...507.1729D,2017ApJ...834..160N, humps}.

\section{Rise time, skewness, and kurtosis parameters}\label{sec:5}

\begin{figure*}
\centering
\includegraphics[width=1\linewidth]{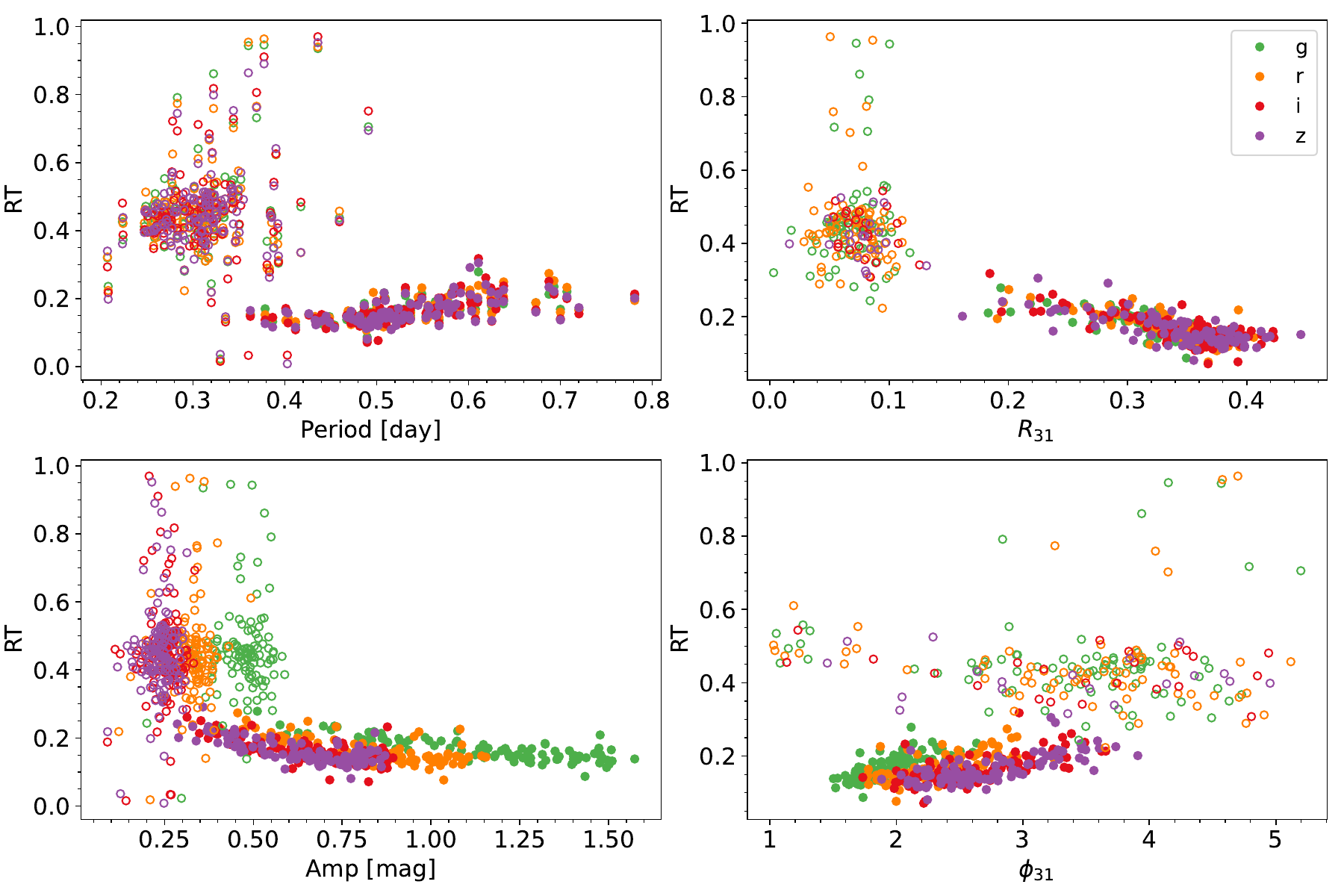}
\caption{The rise time is shown in panel (a) plotted against the period. Panel (b) displays $R_{31}$, panel (c) represents the total amplitude, and panel (d) depicts $\phi_{31}$ for RRab and RRc stars. The $griz$ bands are represented with the same symbols and color scheme as in Figure ~\ref{fig: griz12}.}\label{fig: rt1}
\end{figure*}

The rise time, defined as $ RT = \phi_{\text{max}} - \phi_{\text{min}}$, provides information about the time it takes for a variable star's brightness to increase from its minimum $(\phi_{\text{min}})$ to its maximum $(\phi_{\text{max}})$ light. Figure~\ref{fig: rt1} shows a notable trend in the $griz$ bands for RRab stars, where the rise time generally increases with the period \citep{1981ApJS...46...41S,1984AJ.....89..231C,2005AJ....129..267C}. Additionally, a clear correlation is observed: as the value of $R_{31}$ increases, the rise time tends to decrease. These findings are consistent with previous studies \citep{2004AJ....128..858S,nonblazhko}.
Moreover, our analysis reveals that the rise time increases with larger values of $\phi_{31}$ and decreases with larger total amplitudes. As mentioned above, this is expected, considering that the amplitudes of the RRLs decrease as one moves towards redder bands. In Figure~\ref{fig:PearsonRRab}, these trends are quantified by means of their respective Pearson correlation coefficients, where $RT$ shows negative correlations with amplitude, $R_{21}$, $R_{31}$, and kurtosis across all bands for RRab stars.
On the other hand, for RRc stars significant correlations are not apparent, with rise times largely clumping around a small range of values for the quantities in both axes. The exception is $\phi_{31}$, which covers a large range of values, with only a hint of an anticorrelation with $RT$.

On the other hand, skewness and kurtosis are important statistical measures used in the study of RRLs. These measures provide valuable insights into the shape and asymmetry of the light curves exhibited by these pulsating variable stars \citep{1986ApJ...306..183S,1987ApJ...314..252S,2018A&A...618A..30H}. Skewness is used to quantify the asymmetry of light curves \citep{1986ApJ...306..183S,1987ApJ...314..252S,2018A&A...618A..30H,2021AJ....162..209A}. Positive skewness indicates that the distribution is skewed to the right, while negative skewness indicates a left-skewed distribution. It can also help distinguish between RRLs, such as RRab and RRc, which exhibit different asymmetry patterns in their light curves. Second, high kurtosis indicates a more peaked light curve shape, while low kurtosis indicates a flatter one \citep{2006MNRAS.368.1757W,2018A&A...618A..30H}.

\begin{figure*}[ht]
\centering
\includegraphics[width=1\linewidth]{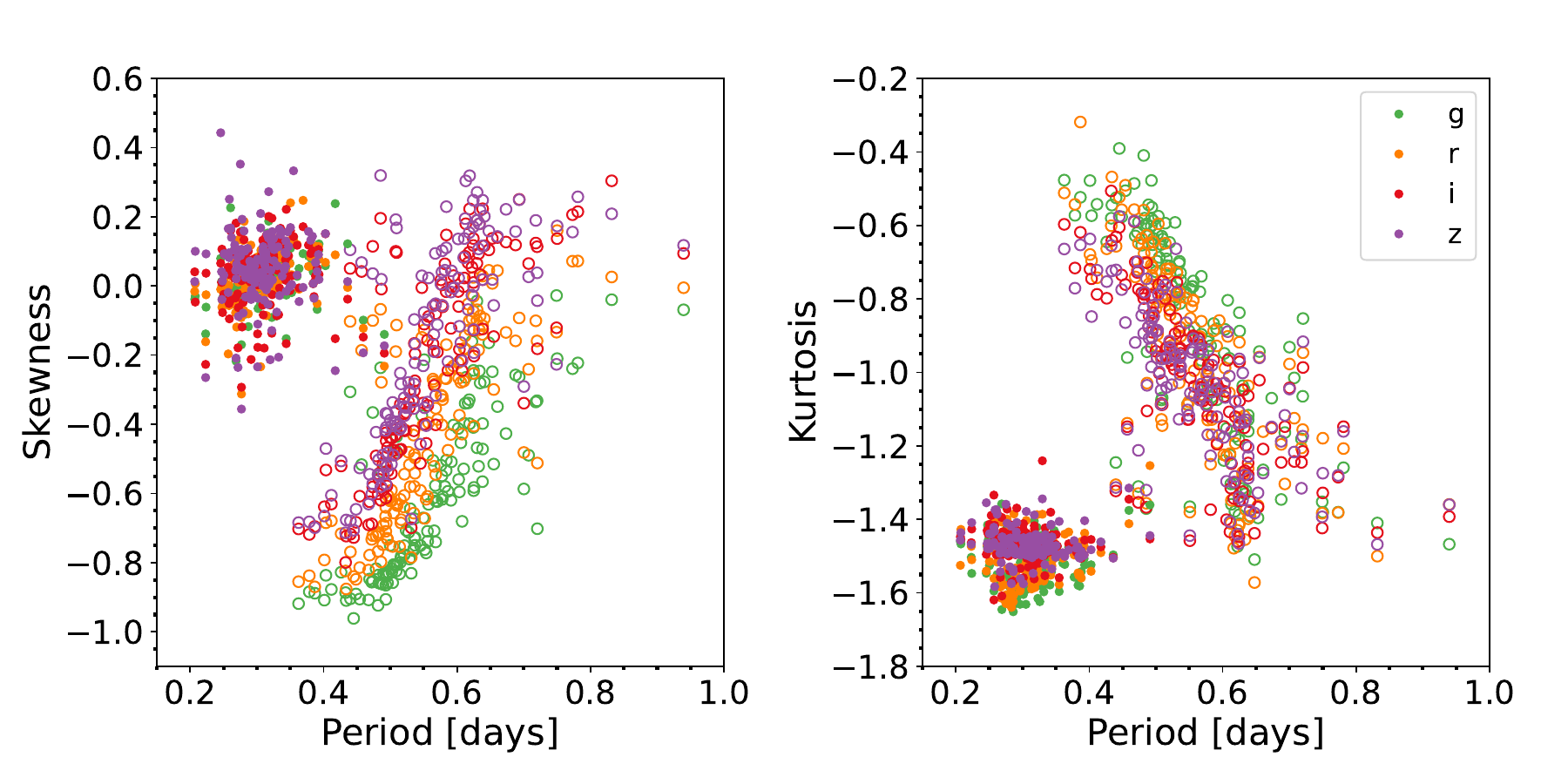}
\caption{Skewness and kurtosis value versus period plot for different photometric bands. The closed circles represent RRc stars and the open circles RRab stars. The $griz$ bands are represented with the same color code as Figure ~\ref{fig: ogle0}.}\label{fig:s}
\end{figure*}

Figure~\ref{fig:s} shows the skewness and kurtosis values for RR Lyrae stars in the $griz$ bands. The left panel gives an overview of the distribution of the skewness values for each band. A very similar behavior can be observed for the skewness values for RRc stars, independent of the band. Note that these values are concentrated around an interval between -0.15 and 0.2. In contrast, for RRab stars, the skewness value tends to increase with period, showing a strong positive correlation with both period and $RT$ (see Figure~\ref{fig:PearsonRRab}). Additionally, there appears to be a strong negative correlation between skewness and amplitude, which also shows a notable dependence on effective wavelength. Similarly, the right panel provides an overview of the kurtosis distribution for each band. The kurtosis values for RRc stars are generally concentrated around -1.6 and -1.4, while for RRab stars, they decrease as the period increases. Based on the Pearson correlation coefficient, we observe a positive correlation between kurtosis and amplitude, as well as $R_{31}$, for RRab stars, and a negative correlation between kurtosis and these same parameters for RRc stars.

\section{Summary}\label{sec:6}
In this work, we studied 1033 previously known RRL stars in the DCP-E DDF, located towards the Galactic bulge in the vicinity of Baade's Window. Extensive multi-band time-series data in the SDSS system ($griz$ bands), both from the main DECAT program and our own observations, were used in this work. We have used these data to provide an extensive set of $griz$ light curve templates for both the RRab and RRc subtypes.  

Our final set of templates, comprised of 136 RRab and 144 carefully selected RRc stars covering a wide range of periods and amplitudes, was derived using Fourier decomposition, performed in an iterative way using a novel outlier removal technique. Compared to traditional methods, our approach helps ensure that good measurements located in the vicinity of the steep rising branches of RRab stars are not inadvertently removed when performing the fitting. 

Using our extensive template set, we analyze the dependence of the multi-band Fourier coefficients, as well as the light curve rise time, skewness, and kurtosis, on parameters such as the period, light curve amplitude, and effective wavelength. Our results, which are provided in the form of extensive tabulations and computer routines, are expected to be especially useful to help detect and characterize RRLs in future surveys such as the upcoming Rubin Observatory's LSST. 

\section{Data availability}
The Table~\ref{table:properties} is only available in electronic form at the CDS via anonymous ftp to cdsarc.u-strasbg.fr (130.79.128.5) or via http://cdsweb.u-strasbg.fr/cgi-bin/qcat?J/A+A/.

\begin{acknowledgements}
Support for this project is provided by ANID's FONDECYT Regular grants \#1171273 and 1231637; ANID's Millennium Science Initiative through grants ICN12\textunderscore 009 and AIM23-0001, awarded to the Millennium Institute of Astrophysics (MAS); and ANID's Basal project FB210003.

C.E.F.L. is additionally supported by DIUDA 88231R11 and LSST Discover Alliance.

C.E.M.-V. is supported by the international Gemini Observatory, a program of NSF NOIRLab, which is managed by the Association of Universities for Research in Astronomy (AURA) under a cooperative agreement with the U.S. National Science Foundation, on behalf of the Gemini partnership of Argentina, Brazil, Canada, Chile, the Republic of Korea, and the United States of America.

This project used data obtained with the Dark Energy Camera (DECam), which was constructed by the Dark Energy Survey (DES) collaboration. Funding for the DES Projects has been provided by the US Department of Energy, the U.S. National Science Foundation, the Ministry of Science and Education of Spain, the Science and Technology Facilities Council of the United Kingdom, the Higher Education Funding Council for England, the National Center for Supercomputing Applications at the University of Illinois at Urbana-Champaign, the Kavli Institute for Cosmological Physics at the University of Chicago, Center for Cosmology and Astro-Particle Physics at the Ohio State University, the Mitchell Institute for Fundamental Physics and Astronomy at Texas A\&M University, Financiadora de Estudos e Projetos, Fundação Carlos Chagas Filho de Amparo à Pesquisa do Estado do Rio de Janeiro, Conselho Nacional de Desenvolvimento Científico e Tecnológico and the Ministério da Ciência, Tecnologia e Inovação, the Deutsche Forschungsgemeinschaft and the Collaborating Institutions in the Dark Energy Survey.

The Collaborating Institutions are Argonne National Laboratory, the University of California at Santa Cruz, the University of Cambridge, Centro de Investigaciones Energéticas, Medioambientales y Tecnológicas–Madrid, the University of Chicago, University College London, the DES-Brazil Consortium, the University of Edinburgh, the Eidgenössische Technische Hochschule (ETH) Zürich, Fermi National Accelerator Laboratory, the University of Illinois at Urbana-Champaign, the Institut de Ciències de l’Espai (IEEC/CSIC), the Institut de Física d’Altes Energies, Lawrence Berkeley National Laboratory, the Ludwig-Maximilians Universität München and the associated Excellence Cluster Universe, the University of Michigan, NSF NOIRLab, the University of Nottingham, the Ohio State University, the OzDES Membership Consortium, the University of Pennsylvania, the University of Portsmouth, SLAC National Accelerator Laboratory, Stanford University, the University of Sussex, and Texas A\&M University.

Based on observations at NSF Cerro Tololo Inter-American Observatory, NSF NOIRLab (NOIRLab Prop. ID 2021A-0921, PI: Catelan), which is managed by the Association of Universities for Research in Astronomy (AURA) under a cooperative agreement with the U.S. National Science Foundation.

\end{acknowledgements}

\bibliographystyle{aa} 
\bibliography{main}

\begin{thebibliography}{89}
\expandafter\ifx\csname natexlab\endcsname\relax\def\natexlab#1{#1}\fi

\bibitem[{{Alam} {et~al.}(2015){Alam}, {Albareti}, {Allende Prieto}, {Anders}, {Anderson}, {Anderton}, {Andrews}, {Armengaud}, {Aubourg}, {Bailey}, {Basu}, {Bautista}, {Beaton}, {Beers}, {Bender}, {Berlind}, {Beutler}, {Bhardwaj}, {Bird}, {Bizyaev}, {Blake}, {Blanton}, {Blomqvist}, {Bochanski}, {Bolton}, {Bovy}, {Shelden Bradley}, {Brandt}, {Brauer}, {Brinkmann}, {Brown}, {Brownstein}, {Burden}, {Burtin}, {Busca}, {Cai}, {Capozzi}, {Carnero Rosell}, {Carr}, {Carrera}, {Chambers}, {Chaplin}, {Chen}, {Chiappini}, {Chojnowski}, {Chuang}, {Clerc}, {Comparat}, {Covey}, {Croft}, {Cuesta}, {Cunha}, {da Costa}, {Da Rio}, {Davenport}, {Dawson}, {De Lee}, {Delubac}, {Deshpande}, {Dhital}, {Dutra-Ferreira}, {Dwelly}, {Ealet}, {Ebelke}, {Edmondson}, {Eisenstein}, {Ellsworth}, {Elsworth}, {Epstein}, {Eracleous}, {Escoffier}, {Esposito}, {Evans}, {Fan}, {Fern{\'a}ndez-Alvar}, {Feuillet}, {Filiz Ak}, {Finley}, {Finoguenov}, {Flaherty}, {Fleming}, {Font-Ribera}, {Foster}, {Frinchaboy}, {Galbraith-Frew}, {Garc{\'\i}a},
  {Garc{\'\i}a-Hern{\'a}ndez}, {Garc{\'\i}a P{\'e}rez}, {Gaulme}, {Ge}, {G{\'e}nova-Santos}, {Georgakakis}, {Ghezzi}, {Gillespie}, {Girardi}, {Goddard}, {Gontcho}, {Gonz{\'a}lez Hern{\'a}ndez}, {Grebel}, {Green}, {Grieb}, {Grieves}, {Gunn}, {Guo}, {Harding}, {Hasselquist}, {Hawley}, {Hayden}, {Hearty}, {Hekker}, {Ho}, {Hogg}, {Holley-Bockelmann}, {Holtzman}, {Honscheid}, {Huber}, {Huehnerhoff}, {Ivans}, {Jiang}, {Johnson}, {Kinemuchi}, {Kirkby}, {Kitaura}, {Klaene}, {Knapp}, {Kneib}, {Koenig}, {Lam}, {Lan}, {Lang}, {Laurent}, {Le Goff}, {Leauthaud}, {Lee}, {Lee}, {Licquia}, {Liu}, {Long}, {L{\'o}pez-Corredoira}, {Lorenzo-Oliveira}, {Lucatello}, {Lundgren}, {Lupton}, {Mack}, {Mahadevan}, {Maia}, {Majewski}, {Malanushenko}, {Malanushenko}, {Manchado}, {Manera}, {Mao}, {Maraston}, {Marchwinski}, {Margala}, {Martell}, {Martig}, {Masters}, {Mathur}, {McBride}, {McGehee}, {McGreer}, {McMahon}, {M{\'e}nard}, {Menzel}, {Merloni}, {M{\'e}sz{\'a}ros}, {Miller}, {Miralda-Escud{\'e}}, {Miyatake}, {Montero-Dorta}, {More},
  {Morganson}, {Morice-Atkinson}, {Morrison}, {Mosser}, {Muna}, {Myers}, {Nandra}, {Newman}, {Neyrinck}, {Nguyen}, {Nichol}, {Nidever}, {Noterdaeme}, {Nuza}, {O'Connell}, {O'Connell}, {O'Connell}, {Ogando}, {Olmstead}, {Oravetz}, {Oravetz}, {Osumi}, {Owen}, {Padgett}, {Padmanabhan}, {Paegert}, {Palanque-Delabrouille}, {Pan}, {Parejko}, {P{\^a}ris}, {Park}, {Pattarakijwanich}, {Pellejero-Ibanez}, {Pepper}, {Percival}, {P{\'e}rez-Fournon}, {P{\'e}rez-R{\`a}fols}, {Petitjean}, {Pieri}, {Pinsonneault}, {Porto de Mello}, {Prada}, {Prakash}, {Price-Whelan}, {Protopapas}, {Raddick}, {Rahman}, {Reid}, {Rich}, {Rix}, {Robin}, {Rockosi}, {Rodrigues}, {Rodr{\'\i}guez-Torres}, {Roe}, {Ross}, {Ross}, {Rossi}, {Ruan}, {Rubi{\~n}o-Mart{\'\i}n}, {Rykoff}, {Salazar-Albornoz}, {Salvato}, {Samushia}, {S{\'a}nchez}, {Santiago}, {Sayres}, {Schiavon}, {Schlegel}, {Schmidt}, {Schneider}, {Schultheis}, {Schwope}, {Sc{\'o}ccola}, {Scott}, {Sellgren}, {Seo}, {Serenelli}, {Shane}, {Shen}, {Shetrone}, {Shu}, {Silva Aguirre}, {Sivarani},
  {Skrutskie}, {Slosar}, {Smith}, {Sobreira}, {Souto}, {Stassun}, {Steinmetz}, {Stello}, {Strauss}, {Streblyanska}, {Suzuki}, {Swanson}, {Tan}, {Tayar}, {Terrien}, {Thakar}, {Thomas}, {Thomas}, {Thompson}, {Tinker}, {Tojeiro}, {Troup}, {Vargas-Maga{\~n}a}, {Vazquez}, {Verde}, {Viel}, {Vogt}, {Wake}, {Wang}, {Weaver}, {Weinberg}, {Weiner}, {White}, {Wilson}, {Wisniewski}, {Wood-Vasey}, {Ye`che}, {York}, {Zakamska}, {Zamora}, {Zasowski}, {Zehavi}, {Zhao}, {Zheng}, {Zhou}, {Zhou}, {Zou}, \& {Zhu}}]{2015ApJS..219...12A}
{Alam}, S., {Albareti}, F.~D., {Allende Prieto}, C., {et~al.} 2015, \apjs, 219, 12

\bibitem[{{Alonso-Garc{\'\i}a} {et~al.}(2012){Alonso-Garc{\'\i}a}, {Mateo}, {Sen}, {Banerjee}, {Catelan}, {Minniti}, \& {von Braun}}]{AG2012}
{Alonso-Garc{\'\i}a}, J., {Mateo}, M., {Sen}, B., {et~al.} 2012, \aj, 143, 70

\bibitem[{{Arp}(1965)}]{Arp1965}
{Arp}, H. 1965, \apj, 141, 43

\bibitem[{{Astropy Collaboration} {et~al.}(2018){Astropy Collaboration}, {Price-Whelan}, {Sip{\H{o}}cz}, {G{\"u}nther}, {Lim}, {Crawford}, {Conseil}, {Shupe}, {Craig}, {Dencheva}, {Ginsburg}, {VanderPlas}, {Bradley}, {P{\'e}rez-Su{\'a}rez}, {de Val-Borro}, {Aldcroft}, {Cruz}, {Robitaille}, {Tollerud}, {Ardelean}, {Babej}, {Bach}, {Bachetti}, {Bakanov}, {Bamford}, {Barentsen}, {Barmby}, {Baumbach}, {Berry}, {Biscani}, {Boquien}, {Bostroem}, {Bouma}, {Brammer}, {Bray}, {Breytenbach}, {Buddelmeijer}, {Burke}, {Calderone}, {Cano Rodr{\'\i}guez}, {Cara}, {Cardoso}, {Cheedella}, {Copin}, {Corrales}, {Crichton}, {D'Avella}, {Deil}, {Depagne}, {Dietrich}, {Donath}, {Droettboom}, {Earl}, {Erben}, {Fabbro}, {Ferreira}, {Finethy}, {Fox}, {Garrison}, {Gibbons}, {Goldstein}, {Gommers}, {Greco}, {Greenfield}, {Groener}, {Grollier}, {Hagen}, {Hirst}, {Homeier}, {Horton}, {Hosseinzadeh}, {Hu}, {Hunkeler}, {Ivezi{\'c}}, {Jain}, {Jenness}, {Kanarek}, {Kendrew}, {Kern}, {Kerzendorf}, {Khvalko}, {King}, {Kirkby}, {Kulkarni},
  {Kumar}, {Lee}, {Lenz}, {Littlefair}, {Ma}, {Macleod}, {Mastropietro}, {McCully}, {Montagnac}, {Morris}, {Mueller}, {Mumford}, {Muna}, {Murphy}, {Nelson}, {Nguyen}, {Ninan}, {N{\"o}the}, {Ogaz}, {Oh}, {Parejko}, {Parley}, {Pascual}, {Patil}, {Patil}, {Plunkett}, {Prochaska}, {Rastogi}, {Reddy Janga}, {Sabater}, {Sakurikar}, {Seifert}, {Sherbert}, {Sherwood-Taylor}, {Shih}, {Sick}, {Silbiger}, {Singanamalla}, {Singer}, {Sladen}, {Sooley}, {Sornarajah}, {Streicher}, {Teuben}, {Thomas}, {Tremblay}, {Turner}, {Terr{\'o}n}, {van Kerkwijk}, {de la Vega}, {Watkins}, {Weaver}, {Whitmore}, {Woillez}, {Zabalza}, \& {Astropy Contributors}}]{price2018astropy}
{Astropy Collaboration}, {Price-Whelan}, A.~M., {Sip{\H{o}}cz}, B.~M., {et~al.} 2018, \aj, 156, 123

\bibitem[{{Astropy Collaboration} {et~al.}(2013){Astropy Collaboration}, {Robitaille}, {Tollerud}, {Greenfield}, {Droettboom}, {Bray}, {Aldcroft}, {Davis}, {Ginsburg}, {Price-Whelan}, {Kerzendorf}, {Conley}, {Crighton}, {Barbary}, {Muna}, {Ferguson}, {Grollier}, {Parikh}, {Nair}, {Unther}, {Deil}, {Woillez}, {Conseil}, {Kramer}, {Turner}, {Singer}, {Fox}, {Weaver}, {Zabalza}, {Edwards}, {Azalee Bostroem}, {Burke}, {Casey}, {Crawford}, {Dencheva}, {Ely}, {Jenness}, {Labrie}, {Lim}, {Pierfederici}, {Pontzen}, {Ptak}, {Refsdal}, {Servillat}, \& {Streicher}}]{2013A&A...558A..33A}
{Astropy Collaboration}, {Robitaille}, T.~P., {Tollerud}, E.~J., {et~al.} 2013, \aap, 558, A33

\bibitem[{{Audenaert} {et~al.}(2021){Audenaert}, {Kuszlewicz}, {Handberg}, {Tkachenko}, {Armstrong}, {Hon}, {Kgoadi}, {Lund}, {Bell}, {Bugnet}, {Bowman}, {Johnston}, {Garc{\'\i}a}, {Stello}, {Moln{\'a}r}, {Plachy}, {Buzasi}, {Aerts}, \& {T'DA collaboration}}]{2021AJ....162..209A}
{Audenaert}, J., {Kuszlewicz}, J.~S., {Handberg}, R., {et~al.} 2021, \aj, 162, 209

\bibitem[{{Baade}(1951)}]{Baade1951}
{Baade}, W. 1951, Publications of Michigan Observatory, 10, 7

\bibitem[{Baart(1982)}]{10.1093/imanum/2.2.241}
Baart, M.~L. 1982, IMA Journal of Numerical Analysis, 2, 241

\bibitem[{{Bla{\v{z}}ko}(1907)}]{Blazhko1907}
{Bla{\v{z}}ko}, S. 1907, Astronomische Nachrichten, 175, 325

\bibitem[{{Boettcher} {et~al.}(2013){Boettcher}, {Willman}, {Fadely}, {Strader}, {Baker}, {Hopkins}, {Tasnim Ananna}, {Cunningham}, {Douglas}, {Gilbert}, {Preston}, \& {Sturner}}]{2013AJ....146...94B}
{Boettcher}, E., {Willman}, B., {Fadely}, R., {et~al.} 2013, \aj, 146, 94

\bibitem[{{Bond} {et~al.}(2004){Bond}, {Udalski}, {Jaroszy{\'n}ski}, {Rattenbury}, {Paczy{\'n}ski}, {Soszy{\'n}ski}, {Wyrzykowski}, {Szyma{\'n}ski}, {Kubiak}, {Szewczyk}, {{\.Z}ebru{\'n}}, {Pietrzy{\'n}ski}, {Abe}, {Bennett}, {Eguchi}, {Furuta}, {Hearnshaw}, {Kamiya}, {Kilmartin}, {Kurata}, {Masuda}, {Matsubara}, {Muraki}, {Noda}, {Okajima}, {Sako}, {Sekiguchi}, {Sullivan}, {Sumi}, {Tristram}, {Yanagisawa}, {Yock}, \& {OGLE Collaboration}}]{Bond2004}
{Bond}, I.~A., {Udalski}, A., {Jaroszy{\'n}ski}, M., {et~al.} 2004, \apjl, 606, L155

\bibitem[{{Braga} {et~al.}(2015){Braga}, {Dall'Ora}, {Bono}, {Stetson}, {Ferraro}, {Iannicola}, {Marengo}, {Neeley}, {Persson}, {Buonanno}, {Coppola}, {Freedman}, {Madore}, {Marconi}, {Matsunaga}, {Monson}, {Rich}, {Scowcroft}, \& {Seibert}}]{Braga2015}
{Braga}, V.~F., {Dall'Ora}, M., {Bono}, G., {et~al.} 2015, \apj, 799, 165

\bibitem[{{Braga} {et~al.}(2019){Braga}, {Stetson}, {Bono}, {Dall'Ora}, {Ferraro}, {Fiorentino}, {Iannicola}, {Inno}, {Marengo}, {Neeley}, {Beaton}, {Buonanno}, {Calamida}, {Contreras Ramos}, {Chaboyer}, {Fabrizio}, {Freedman}, {Gilligan}, {Johnston}, {Lub}, {Madore}, {Magurno}, {Marconi}, {Marinoni}, {Marrese}, {Mateo}, {Matsunaga}, {Minniti}, {Monson}, {Monelli}, {Nonino}, {Persson}, {Pietrinferni}, {Sneden}, {Storm}, {Walker}, {Valenti}, \& {Zoccali}}]{2019A&A...625A...1B}
{Braga}, V.~F., {Stetson}, P.~B., {Bono}, G., {et~al.} 2019, \aap, 625, A1

\bibitem[{{Cacciari}(1984)}]{1984AJ.....89..231C}
{Cacciari}, C. 1984, \aj, 89, 231

\bibitem[{{Cacciari} {et~al.}(2005){Cacciari}, {Corwin}, \& {Carney}}]{2005AJ....129..267C}
{Cacciari}, C., {Corwin}, T.~M., \& {Carney}, B.~W. 2005, \aj, 129, 267

\bibitem[{{Catelan}(2004{\natexlab{a}})}]{2004ASPC..310..113C}
{Catelan}, M. 2004{\natexlab{a}}, in Astronomical Society of the Pacific Conference Series, Vol. 310, IAU Colloq. 193: Variable Stars in the Local Group, ed. D.~W. {Kurtz} \& K.~R. {Pollard}, 113

\bibitem[{{Catelan}(2004{\natexlab{b}})}]{Marcio2004}
{Catelan}, M. 2004{\natexlab{b}}, \apj, 600, 409

\bibitem[{{Catelan}(2009)}]{Catelan2009}
{Catelan}, M. 2009, \apss, 320, 261

\bibitem[{{Catelan}(2023)}]{Marcio2023}
{Catelan}, M. 2023, in Memorie della Societa Astronomica Italiana, Vol.~94, 56

\bibitem[{{Catelan} \& {Smith}(2015)}]{PS}
{Catelan}, M. \& {Smith}, H.~A. 2015, {Pulsating Stars (Wiley-VCH, Weinheim)}

\bibitem[{{Chambers} {et~al.}(2016){Chambers}, {Magnier}, {Metcalfe}, {Flewelling}, {Huber}, {Waters}, {Denneau}, {Draper}, {Farrow}, {Finkbeiner}, {Holmberg}, {Koppenhoefer}, {Price}, {Rest}, {Saglia}, {Schlafly}, {Smartt}, {Sweeney}, {Wainscoat}, {Burgett}, {Chastel}, {Grav}, {Heasley}, {Hodapp}, {Jedicke}, {Kaiser}, {Kudritzki}, {Luppino}, {Lupton}, {Monet}, {Morgan}, {Onaka}, {Shiao}, {Stubbs}, {Tonry}, {White}, {Ba{\~n}ados}, {Bell}, {Bender}, {Bernard}, {Boegner}, {Boffi}, {Botticella}, {Calamida}, {Casertano}, {Chen}, {Chen}, {Cole}, {Deacon}, {Frenk}, {Fitzsimmons}, {Gezari}, {Gibbs}, {Goessl}, {Goggia}, {Gourgue}, {Goldman}, {Grant}, {Grebel}, {Hambly}, {Hasinger}, {Heavens}, {Heckman}, {Henderson}, {Henning}, {Holman}, {Hopp}, {Ip}, {Isani}, {Jackson}, {Keyes}, {Koekemoer}, {Kotak}, {Le}, {Liska}, {Long}, {Lucey}, {Liu}, {Martin}, {Masci}, {McLean}, {Mindel}, {Misra}, {Morganson}, {Murphy}, {Obaika}, {Narayan}, {Nieto-Santisteban}, {Norberg}, {Peacock}, {Pier}, {Postman}, {Primak}, {Rae}, {Rai},
  {Riess}, {Riffeser}, {Rix}, {R{\"o}ser}, {Russel}, {Rutz}, {Schilbach}, {Schultz}, {Scolnic}, {Strolger}, {Szalay}, {Seitz}, {Small}, {Smith}, {Soderblom}, {Taylor}, {Thomson}, {Taylor}, {Thakar}, {Thiel}, {Thilker}, {Unger}, {Urata}, {Valenti}, {Wagner}, {Walder}, {Walter}, {Watters}, {Werner}, {Wood-Vasey}, \& {Wyse}}]{2016arXiv161205560C}
{Chambers}, K.~C., {Magnier}, E.~A., {Metcalfe}, N., {et~al.} 2016, arXiv e-prints, arXiv:1612.05560

\bibitem[{{Clement} {et~al.}(2001){Clement}, {Muzzin}, {Dufton}, {Ponnampalam}, {Wang}, {Burford}, {Richardson}, {Rosebery}, {Rowe}, \& {Hogg}}]{2001AJ....122.2587C}
{Clement}, C.~M., {Muzzin}, A., {Dufton}, Q., {et~al.} 2001, \aj, 122, 2587

\bibitem[{{Clementini} {et~al.}(2023){Clementini}, {Ripepi}, {Garofalo}, {Molinaro}, {Muraveva}, {Leccia}, {Rimoldini}, {Holl}, {Jevardat de Fombelle}, {Sartoretti}, {Marchal}, {Audard}, {Nienartowicz}, {Andrae}, {Marconi}, {Szabados}, {Evans}, {Lecoeur-Taibi}, {Mowlavi}, {Musella}, \& {Eyer}}]{2022yCat..36740018C}
{Clementini}, G., {Ripepi}, V., {Garofalo}, A., {et~al.} 2023, \aap, 674, A18

\bibitem[{{Clementini} {et~al.}(2019){Clementini}, {Ripepi}, {Molinaro}, {Garofalo}, {Muraveva}, {Rimoldini}, {Guy}, {Jevardat de Fombelle}, {Nienartowicz}, {Marchal}, {Audard}, {Holl}, {Leccia}, {Marconi}, {Musella}, {Mowlavi}, {Lecoeur-Taibi}, {Eyer}, {De Ridder}, {Regibo}, {Sarro}, {Szabados}, {Evans}, \& {Riello}}]{2019A&A...622A..60C}
{Clementini}, G., {Ripepi}, V., {Molinaro}, R., {et~al.} 2019, \aap, 622, A60

\bibitem[{{Cowperthwaite} {et~al.}(2016){Cowperthwaite}, {Berger}, {Soares-Santos}, {Annis}, {Brout}, {Brown}, {Buckley-Geer}, {Cenko}, {Chen}, {Chornock}, {Diehl}, {Doctor}, {Drlica-Wagner}, {Drout}, {Farr}, {Finley}, {Foley}, {Fong}, {Fox}, {Frieman}, {Garcia-Bellido}, {Gill}, {Gruendl}, {Herner}, {Holz}, {Kasen}, {Kessler}, {Lin}, {Margutti}, {Marriner}, {Matheson}, {Metzger}, {Neilsen}, {Quataert}, {Rest}, {Sako}, {Scolnic}, {Smith}, {Sobreira}, {Strampelli}, {Villar}, {Walker}, {Wester}, {Williams}, {Yanny}, {Abbott}, {Abdalla}, {Allam}, {Armstrong}, {Bechtol}, {Benoit-L{\'e}vy}, {Bertin}, {Brooks}, {Burke}, {Carnero Rosell}, {Carrasco Kind}, {Carretero}, {Castander}, {Cunha}, {D'Andrea}, {da Costa}, {Desai}, {Dietrich}, {Evrard}, {Fausti Neto}, {Fosalba}, {Gerdes}, {Giannantonio}, {Goldstein}, {Gruen}, {Gutierrez}, {Honscheid}, {James}, {Johnson}, {Johnson}, {Krause}, {Kuehn}, {Kuropatkin}, {Lima}, {Maia}, {Marshall}, {Menanteau}, {Miquel}, {Mohr}, {Nichol}, {Nord}, {Ogando}, {Plazas}, {Reil}, {Romer},
  {Sanchez}, {Scarpine}, {Sevilla-Noarbe}, {Smith}, {Suchyta}, {Tarle}, {Thomas}, {Thomas}, {Tucker}, {Weller}, \& {DES Collaboration}}]{2016ApJ...826L..29C}
{Cowperthwaite}, P.~S., {Berger}, E., {Soares-Santos}, M., {et~al.} 2016, \apjl, 826, L29

\bibitem[{{de Grijs} {et~al.}(2024){de Grijs}, {Whitelock}, \& {Catelan}}]{deGrijs2024}
{de Grijs}, R., {Whitelock}, P.~A., \& {Catelan}, M., eds. 2024, IAU Symposium, Vol. 376, {At the crossroads of astrophysics and cosmology: Period-luminosity relations in the 2020s}

\bibitem[{{Deb} \& {Singh}(2009)}]{2009A&A...507.1729D}
{Deb}, S. \& {Singh}, H.~P. 2009, \aap, 507, 1729

\bibitem[{{Debosscher} {et~al.}(2013){Debosscher}, {Aerts}, {Tkachenko}, {Pavlovski}, {Maceroni}, {Kurtz}, {Beck}, {Bloemen}, {Degroote}, {Lombaert}, \& {Southworth}}]{Debosscher2013}
{Debosscher}, J., {Aerts}, C., {Tkachenko}, A., {et~al.} 2013, \aap, 556, A56

\bibitem[{{Feigelson} {et~al.}(2019){Feigelson}, {Bianco}, \& {Bonito}}]{2019arXiv190108009F}
{Feigelson}, E.~D., {Bianco}, F., \& {Bonito}, S. 2019, arXiv e-prints, arXiv:1901.08009

\bibitem[{{Fukugita} {et~al.}(1996){Fukugita}, {Ichikawa}, {Gunn}, {Doi}, {Shimasaku}, \& {Schneider}}]{1996AJ....111.1748F}
{Fukugita}, M., {Ichikawa}, T., {Gunn}, J.~E., {et~al.} 1996, \aj, 111, 1748

\bibitem[{{Graham} {et~al.}(2023){Graham}, {Knop}, {Kennedy}, {Nugent}, {Bellm}, {Catelan}, {Patel}, {Smotherman}, {Soraisam}, {Stetzler}, {Aldoroty}, {Awbrey}, {Baeza-Villagra}, {Bernardinelli}, {Bianco}, {Brout}, {Clarke}, {Clarkson}, {Collett}, {Davenport}, {Fu}, {Gizis}, {Heinze}, {Hu}, {Jha}, {Juri{\'c}}, {Kalmbach}, {Kim}, {Lee}, {Lidman}, {Magee}, {Mart{\'\i}nez-V{\'a}zquez}, {Matheson}, {Narayan}, {Palmese}, {Phillips}, {Rabus}, {Rest}, {Rodr{\'\i}guez-Segovia}, {Street}, {Vivas}, {Wang}, {Wolf}, \& {Yang}}]{2023MNRAS.519.3881G}
{Graham}, M.~L., {Knop}, R.~A., {Kennedy}, T.~D., {et~al.} 2023, \mnras, 519, 3881

\bibitem[{{Graham} {et~al.}(2021){Graham}, {Nugent}, {Goldstein}, {Kim}, {Narayan}, {Soraisam}, {Shen}, {Street}, {F{\"o}rster}, {Bianco}, {Gizis}, {Smartt}, {Smith}, {Matheson}, {Lee}, {Vivas}, {Clarkson}, {Rest}, {Brout}, {Catelan}, {Jha}, {Wang}, {Palmese}, {Moustakas}, {Sullivan}, {Lidman}, {Williams}, {Inserra}, {Bellm}, {Juri{\'c}}, {Phillips}, {Kennedy}, {Bell}, {Rawls}, {Kalmbach}, {Smotherman}, \& {Eggl}}]{Graham2021}
{Graham}, M.~L., {Nugent}, P.~E., {Goldstein}, D.~A., {et~al.} 2021, Transient Name Server AstroNote, 104, 1

\bibitem[{{Hambleton} {et~al.}(2023){Hambleton}, {Bianco}, {Street}, {Bell}, {Buckley}, {Graham}, {Hernitschek}, {Lund}, {Mason}, {Pepper}, {Pr{\v{s}}a}, {Rabus}, {Raiteri}, {Szab{\'o}}, {Szkody}, {Andreoni}, {Antoniucci}, {Balmaverde}, {Bellm}, {Bonito}, {Bono}, {Botticella}, {Brocato}, {Bu{\v{c}}ar Bricman}, {Cappellaro}, {Carnerero}, {Chornock}, {Clarke}, {Cowperthwaite}, {Cucchiara}, {D'Ammando}, {Dage}, {Dall'Ora}, {Davenport}, {de Martino}, {de Somma}, {Di Criscienzo}, {Di Stefano}, {Drout}, {Fabrizio}, {Fiorentino}, {Gandhi}, {Garofalo}, {Giannini}, {Gomboc}, {Greggio}, {Hartigan}, {Hundertmark}, {Johnson}, {Johnson}, {Jurkic}, {Khakpash}, {Leccia}, {Li}, {Magurno}, {Malanchev}, {Marconi}, {Margutti}, {Marinoni}, {Mauron}, {Molinaro}, {M{\"o}ller}, {Moniez}, {Muraveva}, {Musella}, {Ngeow}, {Pastorello}, {Petrecca}, {Piranomonte}, {Ragosta}, {Reguitti}, {Righi}, {Ripepi}, {Rivera Sandoval}, {Stassun}, {Stroh}, {Terreran}, {Trimble}, {Tsapras}, {van Velzen}, {Venuti}, \& {Vink}}]{2022arXiv220804499H}
{Hambleton}, K.~M., {Bianco}, F.~B., {Street}, R., {et~al.} 2023, \pasp, 135, 105002

\bibitem[{{Holl} {et~al.}(2018){Holl}, {Audard}, {Nienartowicz}, {Jevardat de Fombelle}, {Marchal}, {Mowlavi}, {Clementini}, {De Ridder}, {Evans}, {Guy}, {Lanzafame}, {Lebzelter}, {Rimoldini}, {Roelens}, {Zucker}, {Distefano}, {Garofalo}, {Lecoeur-Ta{\"\i}bi}, {Lopez}, {Molinaro}, {Muraveva}, {Panahi}, {Regibo}, {Ripepi}, {Sarro}, {Aerts}, {Anderson}, {Charnas}, {Barblan}, {Blanco-Cuaresma}, {Busso}, {Cuypers}, {De Angeli}, {Glass}, {Grenon}, {Juh{\'a}sz}, {Kochoska}, {Koubsky}, {Lanza}, {Leccia}, {Lorenz}, {Marconi}, {Marschalk{\'o}}, {Mazeh}, {Messina}, {Mignard}, {Moitinho}, {Moln{\'a}r}, {Morgenthaler}, {Musella}, {Ordenovic}, {Ord{\'o}{\~n}ez}, {Pagano}, {Palaversa}, {Pawlak}, {Plachy}, {Pr{\v{s}}a}, {Riello}, {S{\"u}veges}, {Szabados}, {Szegedi-Elek}, {Votruba}, \& {Eyer}}]{2018A&A...618A..30H}
{Holl}, B., {Audard}, M., {Nienartowicz}, K., {et~al.} 2018, \aap, 618, A30

\bibitem[{{Jayasinghe} {et~al.}(2019){Jayasinghe}, {Stanek}, {Kochanek}, {Shappee}, {Holoien}, {Thompson}, {Prieto}, {Dong}, {Pawlak}, {Pejcha}, {Shields}, {Pojmanski}, {Otero}, {Britt}, \& {Will}}]{2019MNRAS.486.1907J}
{Jayasinghe}, T., {Stanek}, K.~Z., {Kochanek}, C.~S., {et~al.} 2019, \mnras, 486, 1907

\bibitem[{{Jurcsik} \& {Kovacs}(1996)}]{1996A&A...312..111J}
{Jurcsik}, J. \& {Kovacs}, G. 1996, \aap, 312, 111

\bibitem[{{Kovacs}(1998)}]{1998ASPC..135...52K}
{Kovacs}, G. 1998, in Astronomical Society of the Pacific Conference Series, Vol. 135, A Half Century of Stellar Pulsation Interpretation, ed. P.~A. {Bradley} \& J.~A. {Guzik}, 52

\bibitem[{{Layden}(1998)}]{1998AJ....115..193L}
{Layden}, A.~C. 1998, \aj, 115, 193

\bibitem[{{Leavitt} \& {Pickering}(1912)}]{1912Leavitt}
{Leavitt}, H.~S. \& {Pickering}, E.~C. 1912, Harvard College Observatory Circular, 173, 1

\bibitem[{{Marconi} {et~al.}(2015){Marconi}, {Coppola}, {Bono}, {Braga}, {Pietrinferni}, {Buonanno}, {Castellani}, {Musella}, {Ripepi}, \& {Stellingwerf}}]{Marconi2015}
{Marconi}, M., {Coppola}, G., {Bono}, G., {et~al.} 2015, \apj, 808, 50

\bibitem[{{Marconi} {et~al.}(2005){Marconi}, {Nordgren}, {Bono}, {Schnider}, \& {Caputo}}]{Marconi2005}
{Marconi}, M., {Nordgren}, T., {Bono}, G., {Schnider}, G., \& {Caputo}, F. 2005, \apjl, 623, L133

\bibitem[{Mateo {et~al.}(2012)Mateo, Saha, Schechter, \& Levinson}]{DoPHOT}
Mateo, M.~L., Saha, A., Schechter, P.~L., \& Levinson, R. 2012, A DoPHOT User's Manual: now in C!, \url{https://github.com/M1TDoPHOT/DoPHOT_C/blob/master/manual4.0/script.pdf}

\bibitem[{{Medina} {et~al.}(2023){Medina}, {Hansen}, {Mu{\~n}oz}, {Grebel}, {Vivas}, {Carlin}, \& {Mart{\'\i}nez-V{\'a}zquez}}]{2023MNRAS.519.5689M}
{Medina}, G.~E., {Hansen}, C.~J., {Mu{\~n}oz}, R.~R., {et~al.} 2023, \mnras, 519, 5689

\bibitem[{{Miceli} {et~al.}(2008){Miceli}, {Rest}, {Stubbs}, {Hawley}, {Cook}, {Magnier}, {Krisciunas}, {Bowell}, \& {Koehn}}]{2008Miceli}
{Miceli}, A., {Rest}, A., {Stubbs}, C.~W., {et~al.} 2008, \apj, 678, 865

\bibitem[{{Miknaitis} {et~al.}(2007){Miknaitis}, {Pignata}, {Rest}, {Wood-Vasey}, {Blondin}, {Challis}, {Smith}, {Stubbs}, {Suntzeff}, {Foley}, {Matheson}, {Tonry}, {Aguilera}, {Blackman}, {Becker}, {Clocchiatti}, {Covarrubias}, {Davis}, {Filippenko}, {Garg}, {Garnavich}, {Hicken}, {Jha}, {Krisciunas}, {Kirshner}, {Leibundgut}, {Li}, {Miceli}, {Narayan}, {Prieto}, {Riess}, {Salvo}, {Schmidt}, {Sollerman}, {Spyromilio}, \& {Zenteno}}]{2007ApJ...666..674M}
{Miknaitis}, G., {Pignata}, G., {Rest}, A., {et~al.} 2007, \apj, 666, 674

\bibitem[{{Moskalik} \& {Poretti}(2003)}]{2003A&A...398..213M}
{Moskalik}, P. \& {Poretti}, E. 2003, \aap, 398, 213

\bibitem[{{Mullen} {et~al.}(2022){Mullen}, {Marengo}, {Mart{\'\i}nez-V{\'a}zquez}, {Bono}, {Braga}, {Chaboyer}, {Crestani}, {Dall'Ora}, {Fabrizio}, {Fiorentino}, {Monelli}, {Neeley}, {Stetson}, \& {Th{\'e}venin}}]{Mullen2022}
{Mullen}, J.~P., {Marengo}, M., {Mart{\'\i}nez-V{\'a}zquez}, C.~E., {et~al.} 2022, \apj, 931, 131

\bibitem[{{Mullen} {et~al.}(2021){Mullen}, {Marengo}, {Mart{\'\i}nez-V{\'a}zquez}, {Neeley}, {Bono}, {Dall'Ora}, {Chaboyer}, {Th{\'e}venin}, {Braga}, {Crestani}, {Fabrizio}, {Fiorentino}, {Gilligan}, {Monelli}, \& {Stetson}}]{Mullen2021}
{Mullen}, J.~P., {Marengo}, M., {Mart{\'\i}nez-V{\'a}zquez}, C.~E., {et~al.} 2021, \apj, 912, 144

\bibitem[{{Ngeow} {et~al.}(2017){Ngeow}, {Kanbur}, {Bhardwaj}, {Schrecengost}, \& {Singh}}]{2017ApJ...834..160N}
{Ngeow}, C.-C., {Kanbur}, S.~M., {Bhardwaj}, A., {Schrecengost}, Z., \& {Singh}, H.~P. 2017, \apj, 834, 160

\bibitem[{Ngeow {et~al.}(2013)Ngeow, Lucchini, Kanbur, Barrett, \& Lin}]{2013arXiv1309.4297N}
Ngeow, C.-C., Lucchini, S., Kanbur, S., Barrett, B., \& Lin, B. 2013, in IEEE International Conference on Space Science and Communication (IconSpace), 7 (arXiv:1309.4297)

\bibitem[{Percy(2007)}]{percy2007}
Percy, J.~R. 2007, Understanding Variable Stars (Cambridge University Press, Cambridge)

\bibitem[{{Petersen}(1986)}]{1986A&A...170...59P}
{Petersen}, J.~O. 1986, \aap, 170, 59

\bibitem[{{Prudil} {et~al.}(2019){Prudil}, {D{\'e}k{\'a}ny}, {Catelan}, {Smolec}, {Grebel}, \& {Skarka}}]{2019Prudil}
{Prudil}, Z., {D{\'e}k{\'a}ny}, I., {Catelan}, M., {et~al.} 2019, \mnras, 484, 4833

\bibitem[{{Prudil} {et~al.}(2020){Prudil}, {D{\'e}k{\'a}ny}, {Smolec}, {Catelan}, {Grebel}, \& {Kunder}}]{humps}
{Prudil}, Z., {D{\'e}k{\'a}ny}, I., {Smolec}, R., {et~al.} 2020, \aap, 635, A66

\bibitem[{{Prudil} \& {Skarka}(2017)}]{blazhko}
{Prudil}, Z. \& {Skarka}, M. 2017, \mnras, 466, 2602

\bibitem[{{Rest} {et~al.}(2014){Rest}, {Scolnic}, {Foley}, {Huber}, {Chornock}, {Narayan}, {Tonry}, {Berger}, {Soderberg}, {Stubbs}, {Riess}, {Kirshner}, {Smartt}, {Schlafly}, {Rodney}, {Botticella}, {Brout}, {Challis}, {Czekala}, {Drout}, {Hudson}, {Kotak}, {Leibler}, {Lunnan}, {Marion}, {McCrum}, {Milisavljevic}, {Pastorello}, {Sanders}, {Smith}, {Stafford}, {Thilker}, {Valenti}, {Wood-Vasey}, {Zheng}, {Burgett}, {Chambers}, {Denneau}, {Draper}, {Flewelling}, {Hodapp}, {Kaiser}, {Kudritzki}, {Magnier}, {Metcalfe}, {Price}, {Sweeney}, {Wainscoat}, \& {Waters}}]{2014ApJ...795...44R}
{Rest}, A., {Scolnic}, D., {Foley}, R.~J., {et~al.} 2014, \apj, 795, 44

\bibitem[{{Rest} {et~al.}(2005){Rest}, {Stubbs}, {Becker}, {Miknaitis}, {Miceli}, {Covarrubias}, {Hawley}, {Smith}, {Suntzeff}, {Olsen}, {Prieto}, {Hiriart}, {Welch}, {Cook}, {Nikolaev}, {Huber}, {Prochtor}, {Clocchiatti}, {Minniti}, {Garg}, {Challis}, {Keller}, \& {Schmidt}}]{2005ApJ...634.1103R}
{Rest}, A., {Stubbs}, C., {Becker}, A.~C., {et~al.} 2005, \apj, 634, 1103

\bibitem[{{Saha} {et~al.}(2019){Saha}, {Vivas}, {Olszewski}, {Smith}, {Olsen}, {Blum}, {Valdes}, {Claver}, {Calamida}, {Walker}, {Matheson}, {Narayan}, {Soraisam}, {Cunha}, {Axelrod}, {Bloom}, {Cenko}, {Frye}, {Juric}, {Kaleida}, {Kunder}, {Miller}, {Nidever}, \& {Ridgway}}]{Saha_2019}
{Saha}, A., {Vivas}, A.~K., {Olszewski}, E.~W., {et~al.} 2019, \apj, 874, 30

\bibitem[{{S{\'a}nchez-S{\'a}ez} {et~al.}(2023){S{\'a}nchez-S{\'a}ez}, {Arredondo}, {Bayo}, {Ar{\'e}valo}, {Bauer}, {Cabrera-Vives}, {Catelan}, {Coppi}, {Est{\'e}vez}, {F{\"o}rster}, {Hern{\'a}ndez-Garc{\'\i}a}, {Huijse}, {Kurtev}, {Lira}, {Mu{\~n}oz Arancibia}, \& {Pignata}}]{ALERCE2023}
{S{\'a}nchez-S{\'a}ez}, P., {Arredondo}, J., {Bayo}, A., {et~al.} 2023, \aap, 675, A195

\bibitem[{{S{\'a}nchez-S{\'a}ez} {et~al.}(2021){S{\'a}nchez-S{\'a}ez}, {Reyes}, {Valenzuela}, {F{\"o}rster}, {Eyheramendy}, {Elorrieta}, {Bauer}, {Cabrera-Vives}, {Est{\'e}vez}, {Catelan}, {Pignata}, {Huijse}, {De Cicco}, {Ar{\'e}valo}, {Carrasco-Davis}, {Abril}, {Kurtev}, {Borissova}, {Arredondo}, {Castillo-Navarrete}, {Rodriguez}, {Ruz-Mieres}, {Moya}, {Sabatini-Gacit{\'u}a}, {Sep{\'u}lveda-Cobo}, \& {Camacho-I{\~n}iguez}}]{ALERCE2021}
{S{\'a}nchez-S{\'a}ez}, P., {Reyes}, I., {Valenzuela}, C., {et~al.} 2021, \aj, 161, 141

\bibitem[{{Sandage}(2004)}]{2004AJ....128..858S}
{Sandage}, A. 2004, \aj, 128, 858

\bibitem[{{Sandage} {et~al.}(1981){Sandage}, {Katem}, \& {Sandage}}]{1981ApJS...46...41S}
{Sandage}, A., {Katem}, B., \& {Sandage}, M. 1981, \apjs, 46, 41

\bibitem[{{Savino} {et~al.}(2020){Savino}, {Koch}, {Prudil}, {Kunder}, \& {Smolec}}]{Savino2020}
{Savino}, A., {Koch}, A., {Prudil}, Z., {Kunder}, A., \& {Smolec}, R. 2020, \aap, 641, A96

\bibitem[{{Schaltenbrand} \& {Tammann}(1971)}]{1971A&AS....4..265S}
{Schaltenbrand}, R. \& {Tammann}, G.~A. 1971, \aaps, 4, 265

\bibitem[{{Schechter} {et~al.}(1993){Schechter}, {Mateo}, \& {Saha}}]{Schechter_1993}
{Schechter}, P.~L., {Mateo}, M., \& {Saha}, A. 1993, \pasp, 105, 1342

\bibitem[{{Schlafly} {et~al.}(2018){Schlafly}, {Green}, {Lang}, {Daylan}, {Finkbeiner}, {Lee}, {Meisner}, {Schlegel}, \& {Valdes}}]{2018ApJS..234...39S}
{Schlafly}, E.~F., {Green}, G.~M., {Lang}, D., {et~al.} 2018, \apjs, 234, 39

\bibitem[{{Sesar} {et~al.}(2017{\natexlab{a}}){Sesar}, {Hernitschek}, {Mitrovi{\'c}}, {Ivezi{\'c}}, {Rix}, {Cohen}, {Bernard}, {Grebel}, {Martin}, {Schlafly}, {Burgett}, {Draper}, {Flewelling}, {Kaiser}, {Kudritzki}, {Magnier}, {Metcalfe}, {Tonry}, \& {Waters}}]{2017AJ....153..204S}
{Sesar}, B., {Hernitschek}, N., {Mitrovi{\'c}}, S., {et~al.} 2017{\natexlab{a}}, \aj, 153, 204

\bibitem[{{Sesar} {et~al.}(2017{\natexlab{b}}){Sesar}, {Hernitschek}, {Mitrovi{\'c}}, {Ivezi{\'c}}, {Rix}, {Cohen}, {Bernard}, {Grebel}, {Martin}, {Schlafly}, {Burgett}, {Draper}, {Flewelling}, {Kaiser}, {Kudritzki}, {Magnier}, {Metcalfe}, {Tonry}, \& {Waters}}]{SesarHernitschek2017}
{Sesar}, B., {Hernitschek}, N., {Mitrovi{\'c}}, S., {et~al.} 2017{\natexlab{b}}, \aj, 153, 204

\bibitem[{{Sesar} {et~al.}(2010){Sesar}, {Ivezi{\'c}}, {Grammer}, {Morgan}, {Becker}, {Juri{\'c}}, {De Lee}, {Annis}, {Beers}, {Fan}, {Lupton}, {Gunn}, {Knapp}, {Jiang}, {Jester}, {Johnston}, \& {Lampeitl}}]{sesar}
{Sesar}, B., {Ivezi{\'c}}, {\v{Z}}., {Grammer}, S.~H., {et~al.} 2010, \apj, 708, 717

\bibitem[{{Shi} {et~al.}(2021){Shi}, {Qian}, {Li}, \& {Liao}}]{Shi2021}
{Shi}, X.-d., {Qian}, S.-b., {Li}, L.-j., \& {Liao}, W.-p. 2021, \mnras, 505, 6166

\bibitem[{{Simon}(1985)}]{1985ApJ...299..723S}
{Simon}, N.~R. 1985, \apj, 299, 723

\bibitem[{{Simon} \& {Lee}(1981)}]{Simon1981}
{Simon}, N.~R. \& {Lee}, A.~S. 1981, \apj, 248, 291

\bibitem[{{Smith}(1995)}]{1995CAS....27.....S}
{Smith}, H.~A. 1995, {RR Lyrae Stars (Cambridge University Press, Cambridge)}

\bibitem[{{Smolec}(2016)}]{Smolec2016}
{Smolec}, R. 2016, in 37th Meeting of the Polish Astronomical Society, ed. A.~{R{\'o}{\.z}a{\'n}ska} \& M.~{Bejger}, Vol.~3, 22--25

\bibitem[{{Soszy{\'n}ski} {et~al.}(2021){Soszy{\'n}ski}, {Pietrukowicz}, {Skowron}, {Udalski}, {Szyma{\'n}ski}, {Skowron}, {Poleski}, {Koz{\l}owski}, {Mr{\'o}z}, {Ulaczyk}, {Rybicki}, {Iwanek}, {Wrona}, \& {Gromadzki}}]{Soszynski2021}
{Soszy{\'n}ski}, I., {Pietrukowicz}, P., {Skowron}, J., {et~al.} 2021, \actaa, 71, 189

\bibitem[{{Soszy{\'n}ski} {et~al.}(2008){Soszy{\'n}ski}, {Poleski}, {Udalski}, {Szyma{\'n}ki}, {Kubiak}, {Pietrzy{\'n}ski}, {Wyrzykowski}, {Szewczyk}, \& {Ulaczyk}}]{Soszynski2008}
{Soszy{\'n}ski}, I., {Poleski}, R., {Udalski}, A., {et~al.} 2008, \actaa, 58, 163

\bibitem[{{Soszy{\'n}ski} {et~al.}(2014){Soszy{\'n}ski}, {Udalski}, {Szyma{\'n}ski}, {Pietrukowicz}, {Mr{\'o}z}, {Skowron}, {Koz{\l}owski}, {Poleski}, {Skowron}, {Pietrzy{\'n}ski}, {Wyrzykowski}, {Ulaczyk}, \& {Kubiak}}]{2014AcA....64..177S}
{Soszy{\'n}ski}, I., {Udalski}, A., {Szyma{\'n}ski}, M.~K., {et~al.} 2014, \actaa, 64, 177

\bibitem[{{Soszy{\'n}ski} {et~al.}(2019){Soszy{\'n}ski}, {Udalski}, {Wrona}, {Szyma{\'n}ski}, {Pietrukowicz}, {Skowron}, {Skowron}, {Poleski}, {Koz{\l}owski}, {Mr{\'o}z}, {Ulaczyk}, {Rybicki}, {Iwanek}, \& {Gromadzki}}]{2019AcA....69..321S}
{Soszy{\'n}ski}, I., {Udalski}, A., {Wrona}, M., {et~al.} 2019, \actaa, 69, 321

\bibitem[{{Stellingwerf} \& {Donohoe}(1986)}]{1986ApJ...306..183S}
{Stellingwerf}, R. \& {Donohoe}, M. 1986, \apj, 306, 183

\bibitem[{{Stellingwerf} \& {Donohoe}(1987)}]{1987ApJ...314..252S}
{Stellingwerf}, R.~F. \& {Donohoe}, M. 1987, \apj, 314, 252

\bibitem[{{Szab{\'o}} {et~al.}(2011){Szab{\'o}}, {Szabados}, {Ngeow}, {Smolec}, {Derekas}, {Moskalik}, {Nuspl}, {Lehmann}, {F{\.z}r{\'e}sz}, {Molenda-{\.Z}akowicz}, {Bryson}, {Henden}, {Kurtz}, {Stello}, {Nemec}, {Benk{\H{o}}}, {Berdnikov}, {Bruntt}, {Evans}, {Gorynya}, {Pastukhova}, {Simcoe}, {Grindlay}, {Los}, {Doane}, {Laycock}, {Mink}, {Champine}, {Sliski}, {Handler}, {Kiss}, {Koll{\'a}th}, {Kov{\'a}cs}, {Christensen-Dalsgaard}, {Kjeldsen}, {Allen}, {Thompson}, \& {van Cleve}}]{Szabo2011}
{Szab{\'o}}, R., {Szabados}, L., {Ngeow}, C.~C., {et~al.} 2011, \mnras, 413, 2709

\bibitem[{{Tonry} {et~al.}(2012){Tonry}, {Stubbs}, {Lykke}, {Doherty}, {Shivvers}, {Burgett}, {Chambers}, {Hodapp}, {Kaiser}, {Kudritzki}, {Magnier}, {Morgan}, {Price}, \& {Wainscoat}}]{2012ApJ...750...99T}
{Tonry}, J.~L., {Stubbs}, C.~W., {Lykke}, K.~R., {et~al.} 2012, \apj, 750, 99

\bibitem[{{Udalski} {et~al.}(2015){Udalski}, {Soszy{\'n}ski}, {Szyma{\'n}ski}, {Pietrzy{\'n}ski}, {Poleski}, {Pietrukowicz}, {Koz{\l}owski}, {Mr{\'o}z}, {Skowron}, {Skowron}, {Wyrzykowski}, {Ulaczyk}, \& {Pawlak}}]{Udalski2015}
{Udalski}, A., {Soszy{\'n}ski}, I., {Szyma{\'n}ski}, M.~K., {et~al.} 2015, \actaa, 65, 341

\bibitem[{{Usher} {et~al.}(2023){Usher}, {Dage}, {Girardi}, {Barmby}, {Bonatto}, {Chies-Santos}, {Clarkson}, {G{\'o}mez Camus}, {Hartmann}, {Ferguson}, {Pieres}, {Prisinzano}, {Rhode}, {Rich}, {Ripepi}, {Santiago}, {Stassun}, {Street}, {Szab{\'o}}, {Venuti}, {Zaggia}, {Canossa}, {Floriano}, {Lopes}, {Miranda}, {Oliveira}, {Reina-Campos}, {Roman-Lopes}, \& {Sobeck}}]{2023arXiv230617333U}
{Usher}, C., {Dage}, K.~C., {Girardi}, L., {et~al.} 2023, \pasp, 135, 074201

\bibitem[{{Vivas} {et~al.}(2017){Vivas}, {Saha}, {Olsen}, {Blum}, {Olszewski}, {Claver}, {Valdes}, {Axelrod}, {Kaleida}, {Kunder}, {Narayan}, {Matheson}, \& {Walker}}]{vivas2017}
{Vivas}, A.~K., {Saha}, A., {Olsen}, K., {et~al.} 2017, \aj, 154, 85

\bibitem[{{Walker}(1989)}]{Walker1989}
{Walker}, A.~R. 1989, \pasp, 101, 570

\bibitem[{{Wils} {et~al.}(2006){Wils}, {Lloyd}, \& {Bernhard}}]{2006MNRAS.368.1757W}
{Wils}, P., {Lloyd}, C., \& {Bernhard}, K. 2006, \mnras, 368, 1757

\bibitem[{York {et~al.}(2000)York, Adelman, John E.~Anderson, Anderson, Annis, Bahcall, Bakken, Barkhouser, Bastian, Berman, Boroski, Bracker, Briegel, Briggs, Brinkmann, Brunner, Burles, Carey, Carr, Castander, Chen, Colestock, Connolly, Crocker, Csabai, Czarapata, Davis, Doi, Dombeck, Eisenstein, Ellman, Elms, Evans, Fan, Federwitz, Fiscelli, Friedman, Frieman, Fukugita, Gillespie, Gunn, Gurbani, de~Haas, Haldeman, Harris, Hayes, Heckman, Hennessy, Hindsley, Holm, Holmgren, hao Huang, Hull, Husby, Ichikawa, Ichikawa, Željko Ivezić, Kent, Kim, Kinney, Klaene, Kleinman, Kleinman, Knapp, Korienek, Kron, Kunszt, Lamb, Lee, Leger, Limmongkol, Lindenmeyer, Long, Loomis, Loveday, Lucinio, Lupton, MacKinnon, Mannery, Mantsch, Margon, McGehee, McKay, Meiksin, Merelli, Monet, Munn, Narayanan, Nash, Neilsen, Neswold, Newberg, Nichol, Nicinski, Nonino, Okada, Okamura, Ostriker, Owen, Pauls, Peoples, Peterson, Petravick, Pier, Pope, Pordes, Prosapio, Rechenmacher, Quinn, Richards, Richmond, Rivetta, Rockosi,
  Ruthmansdorfer, Sandford, Schlegel, Schneider, Sekiguchi, Sergey, Shimasaku, Siegmund, Smee, Smith, Snedden, Stone, Stoughton, Strauss, Stubbs, SubbaRao, Szalay, Szapudi, Szokoly, Thakar, Tremonti, Tucker, Uomoto, Berk, Vogeley, Waddell, i~Wang, Watanabe, Weinberg, Yanny, \& Yasuda}]{York_2000}
York, D.~G., Adelman, J., John E.~Anderson, J., {et~al.} 2000, \aj, 120, 1579

\bibitem[{{Zong} {et~al.}(2023){Zong}, {Fu}, {Wang}, {Cang}, {Wang}, {Ma}, \& {Zong}}]{nonblazhko}
{Zong}, P., {Fu}, J.-N., {Wang}, J., {et~al.} 2023, \apj, 945, 18

\end{thebibliography}

\begin{appendix} 
\onecolumn
\section{RRc templates}\label{sec:apprrc}

\begin{figure*}[h]
\centering
\includegraphics[width=1.0\linewidth]{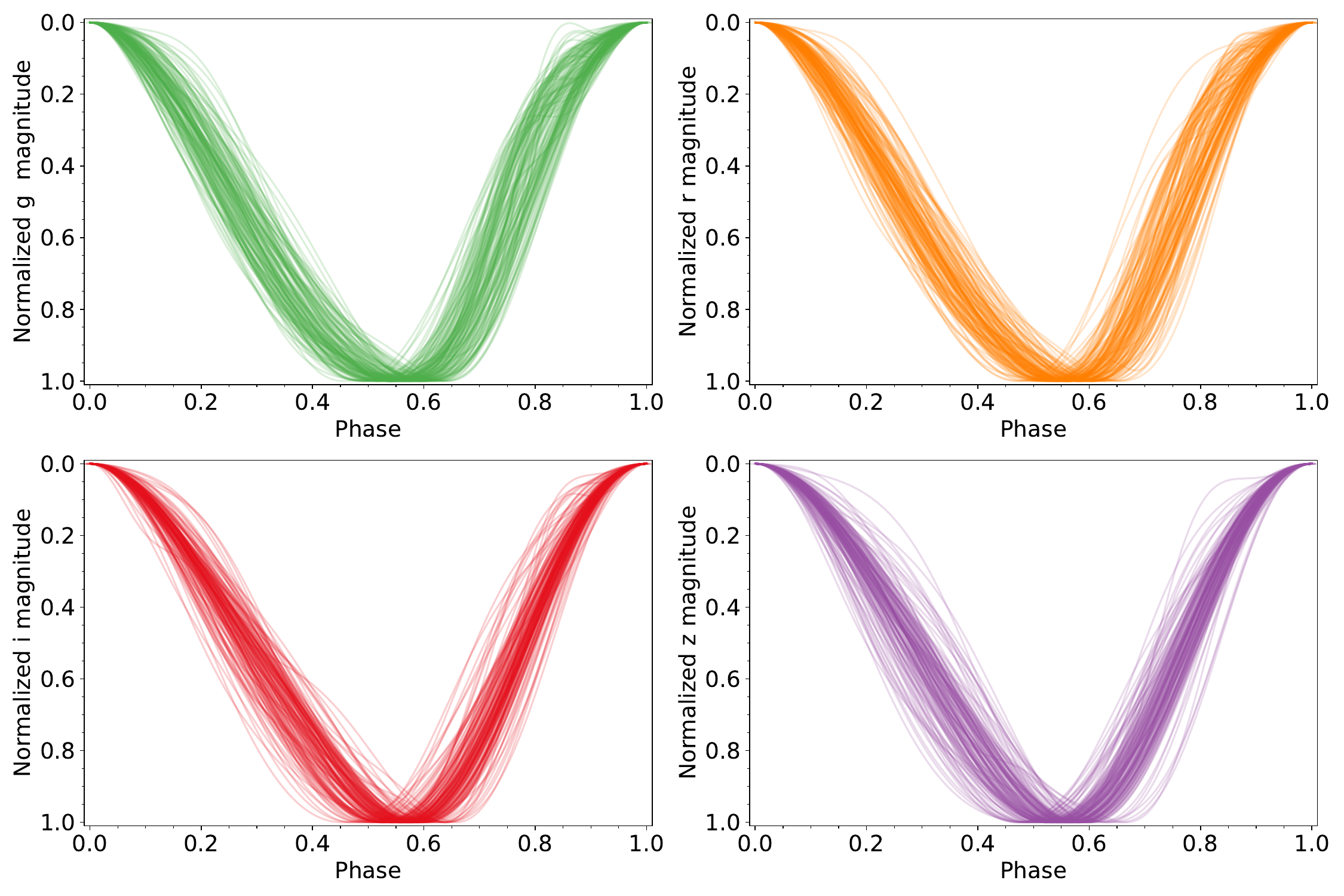}
\caption{Same as Figure~\ref{fig: t0}, but depicting our 144 RRc templates 
in the $griz$ bands.}\label{fig: t1}
\end{figure*}

Figure~\ref{fig: t1} displays the derived templates for the RRc stars in our sample, similarly to what was done in Figure~\ref{fig: t0} in the case of our RRab stars. 

\newpage
\section{Light curve parameters and confusion matrices}\label{sec:confusion}
In Sect.~\ref{sec:fourier}, we described the procedure used to obtain the Fourier coefficients corresponding to our RRab light curve templates. In like vein, Sect.~\ref{sec:5} describes the latter's rise time, skewness, and kurtosis. For each star, the corresponding values, in addition to the periods, amplitudes, Fourier decomposition order, and quality diagnostics (described in Sect.~\ref{sec:templs}) are provided in Table~\ref{table:properties}.

\begin{table*}[h]
\caption{Properties of 136 RRab and 144 RRc stars in the template data sample.}      
\label{table:properties}
\centering
\normalsize
\resizebox{\textwidth}{!}{
\setlength{\tabcolsep}{3pt} 
\begin{tabular}{c c p{1.2cm} c p{1.2cm} p{1.2cm} p{1.2cm} p{1.2cm} p{1.2cm} p{1.2cm} p{1.2cm} p{1.2cm} p{1.2cm} p{1.2cm} p{1.2cm}}   
\hline\hline
ID\tablefootmark{a} & Type & Period\tablefootmark{b} & Band & $A_{\rm tot}$ & $R_{21}$ & $R_{31}$ & $\phi_{21}$ & $\phi_{31}$ & $RT$ & Skewness & Kurtosis & $N$ & $Q$ & $Q^{\prime}$  \\ 
&&(day) & & (mag) &  & & (rad) & (rad) & & & & & & \\
\hline                       
10891 & RRab & 0.492880 & g & 1.313 & 0.477 & 0.358 & 3.829 & 1.719 & 0.120 & -0.832 & -0.602 & 9 &2 &0.710\\
& & & r & 0.953 & 0.504 & 0.393 & 4.076 & 2.094 & 0.118 & -0.665 & -0.708 & 9&2 &0.825\\
& & & i & 0.799 & 0.492 & 0.382 & 4.237 & 2.338 & 0.130 & -0.512 & -0.796 &7&2 & 0.776\\
& & & z & 0.744 & 0.510 & 0.352 & 4.249 & 2.443 & 0.121 & -0.417 & -0.932 & 7&2 &0.724 \\
13817 & RRab & 0.609958 & g & 0.938 & 0.507 & 0.308 & 4.111 & 2.189 & 0.193 & -0.529 & -0.961 & 11 &2 & 0.737\\
& & & r & 0.660 & 0.522 & 0.316 & 4.343 & 2.630 & 0.204 & -0.275 & -1.055 & 11&2 & 0.854\\
& & & i & 0.510 & 0.508 & 0.310 & 4.572 & 3.051 & 0.189 & -0.002 & -1.107& 8 &2 & 0.722\\
& & & z & 0.508 & 0.519 & 0.311 & 4.616 & 3.187 & 0.200 & 0.081 & -1.010 & 8 &2 & 0.711\\
11966 & RRab & 0.507553 & g & 1.181 & 0.498 & 0.356 & 3.865 & 1.819 & 0.150 & -0.807 & -0.652 & 12&2 &0.611\\
& & & r & 0.837 & 0.490 & 0.364 & 4.202 & 2.331 & 0.116 & -0.487 & -0.960 & 12 &1 & 0.533\\
& & & i & 0.645 & 0.414 & 0.371 & 4.107 & 2.274 & 0.140 & -0.431 & -1.087 & 7&2 & 0.579\\
& & & z & 0.647 & 0.462 & 0.357 & 4.199 & 2.404 & 0.152 & -0.396 & -1.048 & 7&2 & 0.607 \\
13354 & RRc & 0.223819 & g & 0.384 & 0.131 & 0.045 & 4.405 & -0.276 & 0.373 & -0.061 & -1.547 & 4&2 &0.821 \\
& & & r & 0.289 & 0.141 & 0.036 & 4.571 & -0.101 & 0.421 & -0.025 & -1.509 & 3&2 & 0.886\\
& & & i & 0.218 & 0.166 & - & 4.784 & - & 0.387 & 0.036 & -1.426 & 2 &2 & 0.735\\
& & & z & 0.218 & 0.183 & - & 4.932 & - & 0.438 & 0.092 & -1.408 & 2&2 & 0.686\\
11557 & RRc & 0.393083 & g & 0.444 & 0.065 & 0.097 & 3.714 & 1.350 & 0.304 & 0.121 & -1.524 & 5&2 & 0.841\\
& & & r & 0.322 & 0.052 & 0.099 & 3.925 & 1.765 & 0.312 & 0.077 & -1.437 & 5&2 & 0.866\\
& & & i & 0.257 & 0.036 & 0.084 & 4.347 & 1.665 & 0.307 & 0.033 & -1.456 & 3&2 & 0.777\\
& & & z & 0.243 & 0.026 & 0.080 & 4.529 & 2.027 & 0.325 & 0.018 & -1.403 & 3&2 & 0.631\\
10729 & RRc & 0.331866 & g & 0.492 & 0.102 & 0.078 & 4.705 & 0.515 & 0.447 & 0.026 & -1.597 & 4&2 &0.689 \\
& & & r & 0.336 & 0.103 & 0.084 & 4.921 & 0.725 & 0.388 & 0.078 & -1.567 & 3&2 & 0.909\\
& & & i & 0.284 & 0.136 & 0.057 & 5.018 & 1.091 & 0.389 & 0.122 & -1.483 & 3&2 &0.801\\
& & & z & 0.276 & 0.121 & - & 5.117 & - & 0.398 & 0.112 & -1.459 & 2&2 &0.748\\
\hline
\end{tabular}}
\tablefoot{
This table is available entirely at the CDS. A portion is provided here as an example to illustrate its structure and content.\\
\tablefoottext{a}{OGLE identification number, in the format OGLE-BLG-RRLYR-{\em ID}.}
\tablefoottext{b}{Period values adopted from OCVS.}
}
\end{table*}

\begin{figure*}[h]
\centering
\includegraphics[width=1\linewidth]{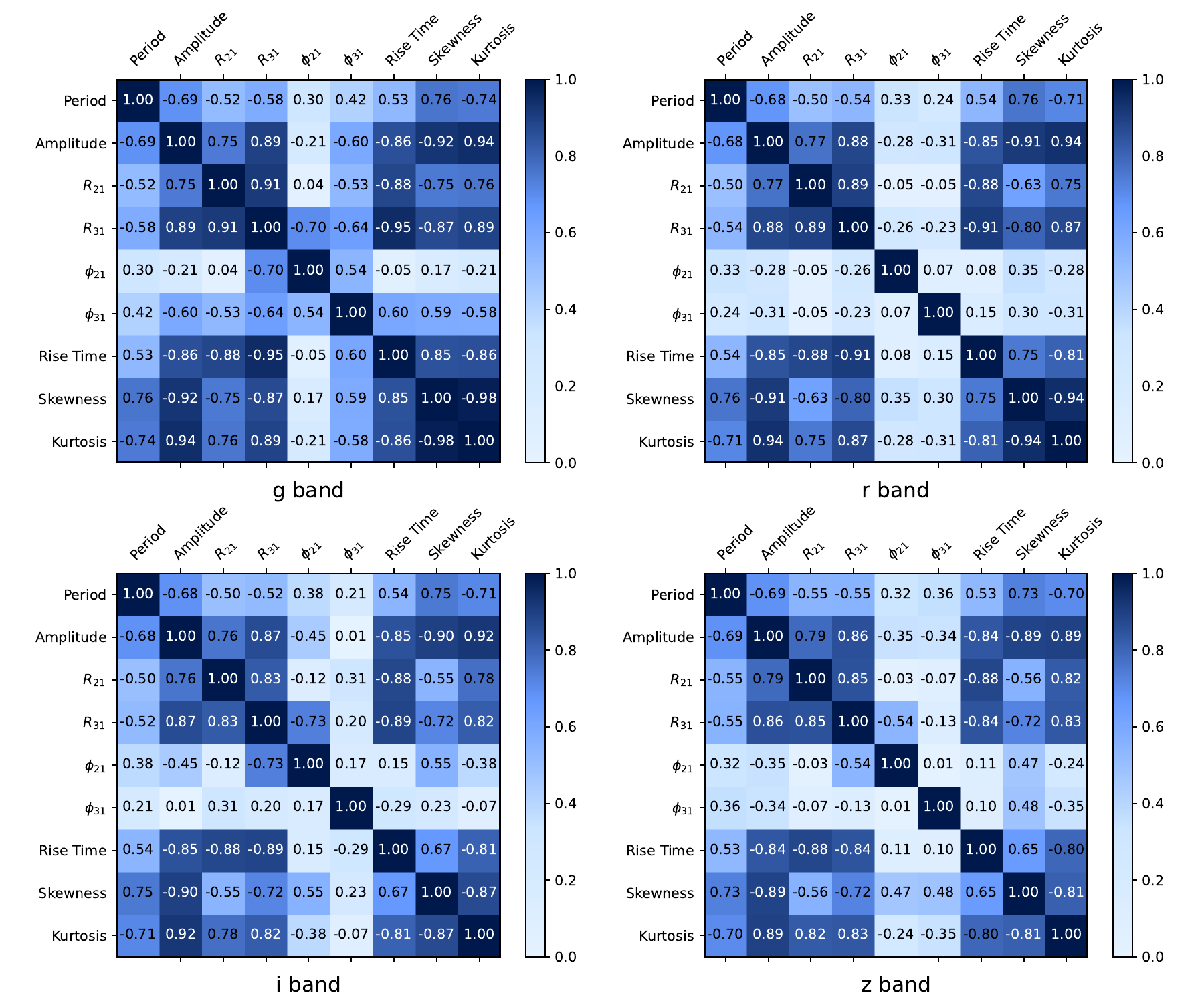}
\caption{Confusion matrices illustrating the Pearson correlation coefficients obtained for the different parameters of RRab stars. Upper left: $g$ band; upper right: $r$ band; bottom left: $i$ band; bottom right: $z$ band.}
\label{fig:PearsonRRab}
\end{figure*}

\begin{figure*}[t]
\centering
\includegraphics[width=1\linewidth]{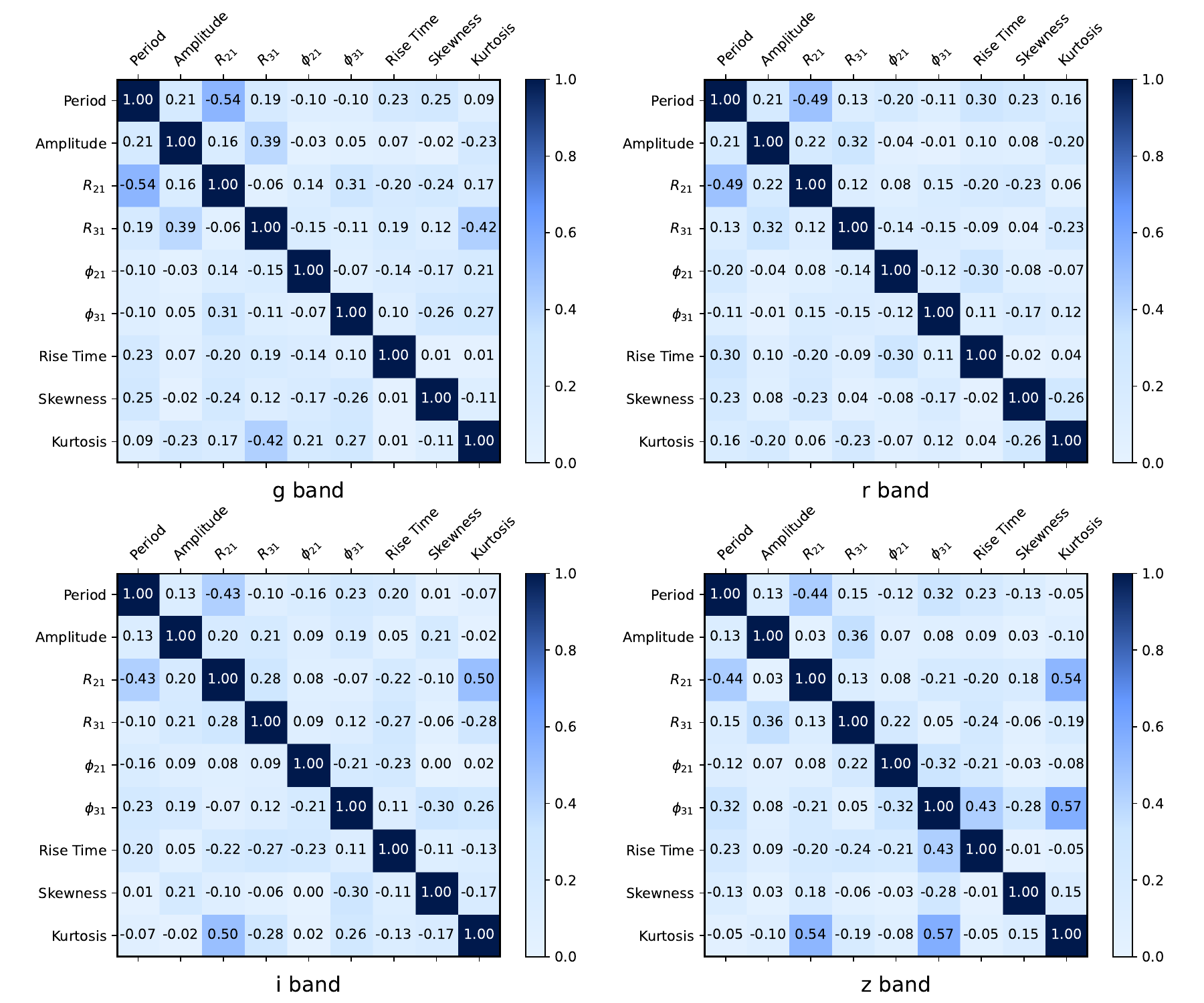}
\caption{Same as Figure~\ref{fig:PearsonRRab}, but for RRc stars.
}\label{fig:PearsonRRc}
\end{figure*}

\clearpage
The interrelation among the aforementioned light curve parameters was quantified by means of their respective Pearson correlation coefficients. The latter are displayed in the form of ``confusion matrices'' for both the RRab (Fig.~\ref{fig:PearsonRRab}) and RRc (Fig.~\ref{fig:PearsonRRc}) stars in our template sample.

\end{appendix}
\end{document}